\documentclass{article}

\usepackage[english]{babel}
\usepackage[letterpaper,top=2cm,bottom=2cm,left=3cm,right=3cm,marginparwidth=1.75cm]{geometry}
\usepackage{amsmath}
\usepackage{amsfonts}
\usepackage{nameref}
\usepackage[sort&compress]{natbib}
\usepackage{xcolor}
\usepackage{enumitem}
\usepackage{graphicx}
\usepackage[colorlinks=true, allcolors=blue]{hyperref}
\usepackage{caption}
\usepackage{subcaption}
\usepackage{float}
\usepackage{comment}
\usepackage[noblocks]{authblk}
\DeclareCaptionLabelFormat{onlycaption}{Caption}
\definecolor{mypink}{RGB}{255,40,250} 

\hypersetup{
    colorlinks=true,
    linkcolor=blue,
    citecolor=mypink,
    urlcolor=mypink
}
\setcitestyle{square,numbers}

\title{\textbf{Tachyonic (In)stability in Randall-Sundrum Braneworld Scenarios}}
\author{Abhirup Karmakar$^{1,}$\footnote{\color{mypink}\textbf{abhirup04karmakar@gmail.com}} \quad and \quad Soumitra SenGupta$^{1,}$\footnote{\color{mypink}\textbf{tpssg@iacs.res.in}} \\ $^{1}$\textit{School of Physical Sciences, Indian Association for the Cultivation of Science, 2A $\&$ 2B, Raja S.C. Mullick Road Kolkata - 700032, India}} 
\date{}
\begin{document}
\maketitle

\vspace{-1cm}
\begin{abstract}
\noindent Low-energy effective theories provide a natural description of four-dimensional physics in higher-dimensional geometries, where the imprint of the bulk geometry appears as parameters of the lower-dimensional theory. Inspired by the Damour-Esposito-Farése (DEF) model of spontaneous scalarization in first generation Scalar-Tensor theories of gravity, we investigate the possibility of tachyonic instability and spontaneous scalarization in braneworld scenarios. We consider the two-brane Randall-Sundrum model where the low-energy effective theory on either brane is of scalar-tensor nature with the extra-dimensional radion playing the role of the scalar. We have determined the possibilities for tachyonic (in)stability of the radion field on either brane in three scenarios: the Randall–Sundrum (RS) model with fine-tuning conditions in which the potential of the radion field vanishes identically, the RS model without fine-tunings where the radion potential arises purely from the gravity sector and the RS model with a bulk stabilizing field that generates a radion potential with a minimum. With the bulk stabilizing field, we have found that on-brane matter with $T>0$ changes the VEV of the radion, destroying the resolution of the gauge hierarchy problem, whereas on-brane matter with $T<0$ does not alter the stability and VEV of the radion. We further determined the exact condition of tachyonic (in)stability of radion field with the on-brane dS$_4$ and AdS$_4$ geometries.
\end{abstract}
\hrulefill \vspace{-0.5cm}
\tableofcontents
\noindent
\hrulefill
\section{Introduction}\label{intro}

The idea of extra spatial dimensions has now gained a well-posed theoretical understanding with our 4D universe viewed as a 3-brane in the 5D bulk spacetime. In various string-inspired models~\cite{1}, the extra spatial dimensions come naturally. However, beyond string theories, the concept of extra spatial dimension has been used in resolving the gauge hierarchy problem of standard model of particle physics. Along with this, it is believed that gravity plays an important role to resolve the issues for beyond standard model physics. As a whole the low-energy extra dimensional models can be classified broadly in two categories - with the large extra dimensional radii~\cite{2} and with the small extra dimensional radii~\cite{3}. In both cases, the extra dimensions are compactified under different topological configurations such that the effective 4D physics emerges from the higher dimensional theory and the imprints of the higher dimensional theory generally gets reflected in the effective 4D theory. In this work, we will mainly focus on the two-brane Randall-Sundrum (RS) warped geometry model~\cite{3}, along with its variants, has been thoroughly studied over the past two decades~\cite{4,5,6,7,8,9,10,11,12,13,14,15,16,17,18,19,20,21,Soham}. In RS model we consider a factorizable 5D geometry with the conventional warp factor which is essential in order to address the problem of gauge hierarchy~\cite{hie,hie2} between the gravitational and electroweak scales by providing an exponential suppression of the mass scale at the visible brane without any additional large fundamental hierarchy.\\

\noindent The separation between the two 3-branes is parametrized by a scalar degree of freedom which comes from the fundamental 5D theory. This is called the \textit{radion} or \textit{modulus} field. Upon dimensional reduction~\cite{22} in the original 5D action, the radion field naturally appears as a dynamical scalar in the 4D effective action and the effective theory becomes scalar-tensor theory on the either branes. One needs to have a stabilization mechanism of the modulus field in order to have a stable effective theory. And this is one of the most challenging task to stabilize the distance between the two branes. Goldberger and Wise~\cite{23,67} proposed a mechanism \textit{in vacuum} to stabilize the radion by introducing a bulk stabilizing field which provides a stable minima, without invoking any unnatural fine-tunings of the parameters of the model. In the absence of on-brane matter, upon the radion stabilization, the radion settles at the desired vacuum expectation value and the on-brane effective theory reduces to that of general relativity (GR). There are studies involving the thermal description of scalar-tensor theories~\cite{25} or more specifically braneworld theories which describes the attractor-to-GR mechanism in both the Planck and TeV brane under different conditions~\cite{26}. However, the presence of on-brane matter may substantially modify the effective radion dynamics raising some important questions whether the vacuum-stabilized configuration always remains stable or not.\\

\noindent In this work, the instabilities arising from the second derivative of the radion potential is studied in details for the 5D RS braneworld scenarios. The second derivative of any effective potential of any dynamical field is defined as the effective mass squared of that field. There is specific kind of instability associated with the negativity of the effective mass squared of any field. These are called \textit{tachyonic instabilities}. Under tachyonic instabilities, small perturbations of the field grows exponentially indicating an unstable chosen background. If there are non-linearity present in the theory, the instability may be quenched and the field evolves towards a new vacuum. A closely related phenomenon is \textit{spontaneous scalarization}~\cite{27} in scalar-tensor theories. Scalarization is a mechanism that endows strongly self-gravitating bodies, such as neutron stars~\cite{37,38,39,40,41,42,43} and black holes~\cite{44,45,46,47,48,49,50}, with a scalar-ﬁeld conﬁguration. It resembles a phase transition in that the scalar conﬁguration appears only when a certain quantity that characterizes the compact object, for example, its compactness or spin, is beyond a threshold. The system should have to be prone to tachyonic instability in order to spontaneous scalarization occur. This mechanism of spontaneous scalarization was proposed by Damour and Esposito-Farése (1993)~\cite{28}. They showed that a specific type of nonminimal coupling between scalar field and gravity (or matter, after a field redefinition) leads to a theory that is indistinguishable from GR in weak-field gravitational experiments and yet predicts order unity deviations from general-relativistic expectations in the strong-gravity regime of neutron stars. In DEF model~\cite{28,29,30,31,32,33}, the effective action of the scalar-tensor theory is generally written in Einstein frame with no potential term. But in literature, there are works with a potential term in the effective action~\cite{34,35,36}. Motivated by the DEF mechanism, the tachyonic (in)stability of the radion field in RS braneworld scenarios is discussed. Since the effective 4D theory is scalar–tensor in nature, with the radion as the scalar degree of freedom, the formalism of spontaneous scalarization can be naturally extended to braneworld models. While the original DEF scenario is concerned with compact objects, corresponding to matter with $T<0$ (non-relativistic matter), we analyze the radion dynamics for all possible matter sectors $T=0, T>0$ and $T<0$. In particular, we study the Tuned-RS, Detuned-RS, and Goldberger–Wise stabilized RS models and discuss the conditions under which spontaneous scalarization, or more appropriately \textit{spontaneous radionization} may occur. Our analysis therefore suggests a higher-dimensional gravitational origin for the scalar degree of freedom responsible for spontaneous scalarization in compact objects. For the sake of completeness, we briefly discuss the tachyonic instabilities in $f(R)$ theories of gravity as $f(R)$ gravity is dual to a specific scalar-tensor theory. We further discuss the tachyonic instability of radion field in dS$_4$ and AdS$_4$ brane geometries.\\

\noindent The paper is organized as follows: the section \ref{2} gives a compact introduction of tachyonic instability in scalar field theory at the leading order scalar perturbation. It is divided into two subsections \ref{flat} and \ref{curved} which gives the concept of tachyonic instability in flat and curved spacetime, respectively. The section \ref{ST} gives a DEF model-inspired description of tachyonic instability conditions and possibility of spontaneous scalarization in scalar-tensor theories in Einstein frame as well as in Jordan frame. And in section \ref{fR}, the formalism of scalar-tensor theory is applied to $f(R)$ theories of gravity for completeness. The section \ref{RS section} is the main attention of our work where the tachyonic instability of radion field is discussed for different scenarios - tuned-RS (section \ref{Tuned-RS}), detuned-RS (section \ref{detuned-RS}) and Goldberger Wise stabilized RS (section \ref{GW}). In section \ref{effective}, the full 4D effective action on RS brane is discussed in both Jordan and Einstein frame which is indispensable in the discussion on braneworld scenarios. The section \ref{flrw} gives a simple example of on-brane dS$_4$ and AdS$_4$ geometries in order to demonstrate the fact that we need to specify the background geometry to get the exact condition for tachyonic instability in gravitational scenario.

\section{What is Tachyonic Instability ?}\label{TI}

\subsection{In Flat Spacetime}\label{flat}

It is very much instructive to review the dynamics of a real scalar field $\phi$ in Minkowski background with a canonical kinetic term and a quartic self-interacting potential. The action of the scalar field $\phi$ is given by
\begin{equation}\label{1}
    S = \int d^4x\left[-\frac{1}{2}\eta^{\mu\nu}\partial_\mu\phi\partial_\nu\phi - V(\phi)\right]
\end{equation}
where $\eta^{\mu\nu}$ is the inverse Minkowski metric and
\begin{equation}\label{2}
    V(\phi) = \frac{1}{2}\mu^2\phi^2 + \frac{1}{4}\lambda\phi^4
\end{equation}
is the potential with $\mu$ being the bare mass of the scalar field and $\lambda$ is called the coupling constant.\\
Varying the action in (\ref{1}) with respect to the scalar field $\phi$, one can obtain the following equation of motion of the scalar field $\phi$
\begin{equation}\label{3}
    \Box_\eta\phi - \mu^2\phi -\lambda\phi^3 = 0
\end{equation}
where $\Box_\eta = \eta^{\mu\nu}\partial_\mu\partial_\nu$ is the flat spacetime d'Alembertian. See $\phi = 0$ is a solution of the above non-linear equation of motion of $\phi$. Consider the small perturbation of the scalar field $\delta\phi$ around $\phi = 0$. At the leading order approximation, we obtain the following linearized equation from eq. (\ref{3})
\begin{equation}\label{4}
    (\Box_\eta - \mu^2)\delta\phi = 0
\end{equation}
The corresponding dispersion relation is $\omega^2 = k^2 + \mu^2$, where $\omega$ is the angular frequency and $k$ is the wavenumber. The relevant solutions to eq. (\ref{4}) are plane waves $\delta\phi \sim \exp\{i(k\cdot x - \omega t)\}$. For small-$k$ modes, if $\mu^2 > 0$, the perturbation $\delta\phi$ oscillates with real frequency $\omega$, whereas if instead $\mu^2 < 0$, the perturbation $\delta\phi$ grows exponentially with time. This is known as the \textit{tachyonic instability} and we say one encounters the tachyonic instability for $\mu^2 < 0$ in flat spacetime.
Although it seems at first glance that the tachyonic instability is catastrophic, but it is not. When perturbation $\delta\phi$ grows exponentially with time, the linearized equation very soon becomes invalid and thus the nonlinear self-interacting term $\lambda\phi^3$ becomes important. Assume $\lambda>0$ and $\mu^2<0$, which is the well-known Mexican-Hat potential. In that case, eq. (\ref{3}) then admits a second solution with constant $\phi$, which we denote as $\phi_{min}$, the minimum of the potential and it is different from the earlier $\phi = 0$. Eq. (\ref{3}) implies the following 
\begin{equation}\label{5}
    \frac{dV}{d\phi} = \mu^2\phi + \lambda\phi^3 = 0 \implies \phi_{min} = \sqrt{\frac{-\mu^2}{\lambda}}
\end{equation}
Thus, the tachyonic instability simply drives the scalar field $\phi$ away from the unstable local maximum ($\phi=0$) of the potential and towards a stable minimum ($\phi = \phi_{min}$). This process is known as the \textit{tachyon condensation}. This is associated to a phase transition of the system. \\

\noindent The important lesson from this section is that the tachyonic instability is nothing but the consequence of choosing the wrong minima (that is, a unstable maximum) for doing perturbation theory. But if we have non-linear interactions present in our theory, in that case the tachyonic instability will eventually be quenched by the non-linear terms driving the field to a stable configuration. Now, we will see in the following subsection, the tachyonic instability of scalar fields in curved backgrounds.

\subsection{In Curved Spacetime}\label{curved}
In previous section, we discussed about the tachyonic instability of a scalar field $\phi$ in flat Minkowski background. The generalization to curved spacetime is simple. We promote the Minkowski metric $\eta_{\mu\nu}$ to some general metric $g_{\mu\nu}$ and therefore the linearized scalar field equation becomes
\begin{equation}\label{6}
    (\Box - \mu^2)\delta\phi = 0
\end{equation}
where $\Box = g^{\mu\nu}\nabla_\mu\nabla_\nu$ is the curved spacetime d'Alembertian with $\nabla_\mu$ the covariant derivative.
The key distinction between flat and curved spacetime is that $\mu^2<0$ is no longer a sufficient condition for having a tachyonic instability in curved spacetime. \\

\noindent One simple example to demonstrate the above statement is the study of tachyonic instability of the scalar field in Schwarzschild background. The metric $g_{\mu\nu}$ of the Schwarzschild spacetime is given by the following line element (considering $G=1$ for this case)
\begin{equation}
    ds^2 = -\left(1 - \frac{2M}{r}\right)dt^2 + \frac{dr^2}{\left(1 - \frac{2M}{r}\right)} +r^2d\Omega_2^2
\end{equation}
where, $d\Omega_2^2 = d\theta^2 + \sin^2\theta d\phi^2$ is the line element of unit-radius $S^2$ and $M$ is the mass of the compact object. It is easy to study as Schwarzschild geometry is spherically symmetric and one can expand the scalar perturbation $\delta\phi$ in spherical harmonics basis $Y_{lm}(\theta,\phi)$ with a harmonic time-dependence
\begin{equation}
    \delta\phi = \sum_{l,m}\frac{\psi_{lm}(r)}{r}Y_{lm}(\theta,\phi)e^{-i\omega t}
\end{equation}
and obtain a Schr\"odinger-like equation (the Regge–Wheeler equation) from eq.(\ref{6}) as follows
\begin{equation}
    \frac{d^2\psi_{lm}}{dr_*^2} + \left[\omega^2 - V_{eff}(r)\right]\psi_{lm} = 0
\end{equation}
where $r_*$ is the tortoise coordinate defined as $dr/dr_* = 1-2M/r$ and $V_{eff}$ is an effective potential given by
\begin{equation}
    V_{eff}(r) = \left(1 - \frac{2M}{r}\right)\left[\frac{l(l+1)}{r^2}+\frac{2M}{r^3}+\mu^2\right]
\end{equation}
which encodes information about the curved background geometry. The sufficient condition (not necessary) to have an tachyonic instability in Schwarzschild geometry is
\begin{equation}\label{7}
    \int_{-\infty}^\infty dr_*V_{eff}(r) \leq 0
\end{equation}
where $r_* = -\infty$ corresponds to the horizon radius $r=2M$ in usual coordinates. $M=0$ corresponds to the flat spacetime and hence $\mu^2 < 0$ will lead to the tachyonic instability. But for $M\neq0$, the situation is different and only $\mu^2 < 0$ does not guaranty the tachyonic instability. For the specific case of Schwarzschild, $\mu^2$ must be sufficiently negative in order to have a tachyonic instability. \\

\noindent The main lesson from this section is that the tachyonic instability condition depends on the background geometry. And the threshold for tachyonic instability may change for different spacetimes. We will more see some example of de-Sitter and Anti de-Sitter spacetimes after discussing Randall-Sundrum scenarios in section \ref{flrw}.\\

\noindent In the next section \ref{sec3}, we will consider a broad class of gravity theories which contains an additional scalar degree freedom. These theories are called \textit{scalar-tensor theories}. Such low-energy effective modified gravity theories can arise from different fundamental origins. One of the simplest origin may be the higher curvature corrections present in the standard Einstein-Hilbert action, called $f(R)$ theories of gravity which inherently contains a gravitational scalar degree of freedom and these are introduced in the next section itself in a dedicated subsection. There is also another interesting origin of scalar-tensor theories, which are higher dimensional gravity theories, like a 5D warped braneworld models. This type of theory also contains a 4D effective scalar degree of freedom, known as \textit{radion}, which parametrizes fluctuations of the extra-dimensional modulus, in the 4D effective action (on-brane action) that originates from the 5D fundamental theory. This essentially forms the basis of our main work presented in the section \ref{RS section}. In the final concluding section \ref{conclusion}, we further discuss that there may be a non-trivial duality between the on-brane radion degree of freedom and $f(R)$ scalar degree of freedom at the inflationary regime.

\section{(In)stability in Scalar-Tensor and $f(R)$ Theories}\label{sec3}

\subsection{Scalar-Tensor Theories}\label{ST}

Consider a first-generation scalar-tensor theory action in Einstein frame with $\varphi$ being the real Einstein frame scalar field with a canonical kinetic term and a potential~\cite{27,28}
\begin{equation}\label{9}
    S = \frac{1}{16\pi G_*}\int d^4x \sqrt{-g} \left[R - \frac{1}{2}\nabla_\mu\varphi\nabla^\mu\varphi - V(\varphi)\right] + S_M[\psi, \mathcal A^2(\varphi)g_{\mu\nu}]
\end{equation}
where $R$ is the Ricci scalar and $S_M$ is the matter action with $\psi$ denoting the matter fields. The coupling with the matter field $\psi$ is through the function $\mathcal A^2(\varphi)$ and $G_*$ denotes some effective gravitational constant. All the terms in the above action are atleast quadratic in $\varphi$.\\

\noindent Variation of the action with respect to the scalar field $\varphi$ yields the following equation of motion of $\varphi$
\begin{equation}\label{10}
    \Box \varphi = V'(\varphi) - 16\pi G_*\alpha(\varphi)T \equiv V_{eff}'(\varphi)
\end{equation}
where $V'(\varphi) = dV/d\varphi$, $V_{eff}(\varphi)$ is the effective potential of the Einstein frame scalar field $\varphi$ in the presence of matter and
\begin{equation}\label{11}
    \alpha(\varphi) = \frac{d\ln\mathcal A(\varphi)}{d\varphi}
\end{equation}
See $T = g_{\mu\nu}T^{\mu\nu}$ is the trace of the matter energy-momentum tensor in the Einstein frame defined as 
\begin{equation}\label{12}
    \delta S_M \equiv \frac{1}{2}\int d^4x \sqrt{- g}\; T^{\mu\nu}\delta g_{\mu\nu}
\end{equation}
The function $\alpha(\varphi)$ controls the coupling between the scalar field and matter. Varying the action (\ref{9}) with respect to the inverse metric yields the modified field equations (which are second order non-linear partial differential equations) as follows
\begin{equation}\label{13}
    G_{\mu\nu} = 8\pi G_*T_{\mu\nu} + \frac{1}{2}\nabla_\mu\varphi\nabla_\nu\varphi - \frac{1}{2}g_{\mu\nu}\left(\frac{1}{2}\nabla^\sigma\varphi\nabla_\sigma\varphi + V(\varphi)\right)
\end{equation}
If $V'_{eff}(\varphi_0) = 0$ for some constant field configuration $\varphi = \varphi_0$ with $T \neq 0$, then it will be an admissible solution of the theory. In fact, it will be a solution of GR with an effective cosmological constant $\Lambda_{eff} = V(\varphi_0)/2$. Now we perturb eq.(\ref{10}) linearly in $\varphi$ in a fixed background metric that is a solution of GR with an effective cosmological constant and we obtain the following linearized equation
\begin{equation}\label{14}
    (\Box - \mu_{eff}^2)\delta\varphi = 0
\end{equation}
with
\begin{equation}\label{15}
    \mu_{eff}^2 = V''(\varphi_0) - 16\pi G_*\alpha'(\varphi_0)T \equiv V_{eff}''(\varphi_0)
\end{equation}
where $\alpha' = d\alpha/d\varphi$, $\mu_{eff}$ is the effective mass of the scalar perturbation and $V''(\varphi_0)$ is the bare mass of the scalar field. The effective mass squared contains an extra piece of matter term in addition to the bare mass of the field. The eq.(\ref{15}) is our master equation. Depending on the sign of $V''_{eff}(\varphi_0)$, there will be different cases, where there is tachyonic instability ($\mu_{eff}^2 < 0$) or no tachyonic instability ($\mu_{eff}^2 > 0$). As we discussed in previous sections, $\mu_{eff}^2 < 0$ is needed for tachyonic instability to occur, but it is not sufficient. It is dependent on the fixed background geometry. \\

\noindent One of the important consequences of tachyonic instability is spontaneous scalarization, which is a process by which the tachyonic instabilities are quenched by non-linearities (that is, non-linear terms present in the effective potential).
For compact objects - like stars, black holes or neutron stars, generally $T<0$. In Damour-Esposito-Far\'ese (DEF) model, $V(\varphi) = 0$. In the DEF model, for $\alpha'(\phi_0) < 0$, the scalar field can develop a tachyonic instability around the spacetime that is the solution of GR. This instability can be quenched by nonlinearities and that the outcome will be a compact object with a nontrivial field configuration. Thus the compact objects become \textit{scalarized} and this process is called the \textit{spontaneous scalarization}. And these scalarized configuration is not a solution of GR and given by modified field equations in eq.(\ref{13}).\\

\noindent One can do the similar analysis of tachyonic instabilities in Jordan frame instead of Einstein frame. The Einstein and Jordan frame is related by a conformal transformation through $\mathcal A(\varphi)$ function, $\tilde g_{\mu\nu} = \mathcal A^2(\varphi)g_{\mu\nu}$, where $\tilde g,g$ are the metric in Jordan and Einstein frame, respectively. In Einstein frame, the field couples to matter, whereas in Jordan frame, the field couples to curvature. In Jordan frame, the action will contain $\varphi^2 R$ term, which in linearized scalar field equation translates into $\mu_{eff}^2 \sim R$ (considering $V = 0$). We can see this from another point of view. The fixed background of our perturbation is a solution of GR. Hence the trace of energy-momentum tensor of matter is related to the Ricci scalar as $R = -8\pi G_*T$. In DEF model, considering $V(\varphi) = 0$, in Jordan frame the effective mass squared is
\begin{equation}\label{16}
    \mu_{eff}^2 = 2\alpha'(\varphi_0)R
\end{equation}
Thus in Jordan frame, curvature couples with the field and enters in the effective mass in perturbation. Both the Einstein and Jordan frame analysis are equivalent.\\

\noindent In this work, we will only stick to the Einstein frame as the matter couples with the field in this frame and it is useful to study the matter-induced tachyonic instabilities in this frame. If one decides to fix the coupling with matter in a particular frame, then one can no longer change the frame. However, the tachyonic instability analysis can be done equivalently in both Einstein or Jordan frames. The two descriptions - Einstein (couples to matter) and Jordan (couples to curvature) frames essentially capture two equivalent interpretations of the scalarization mechanism and both are related through the trace equation ($R$ and $T$ equation) of the theory.\\

\noindent Below we provide the instability analysis of $f(R)$ theories for the sake of completeness, as $f(R)$ theories are dual to a specific scalar-tensor theory, known as the Brans-Dicke theory with a vanishing Brans-Dicke parameter.

\subsection{$f(R)$ Theories}\label{fR}

The primary motivation behind considering $f(R)$ theories of gravity\footnote{In this paper, we are considering $f(R)$ theories in metric formalism only.} comes from loop corrections to the matter fields in curved spacetime~\cite{68,69} and their potential in explaining early-time inflation and late-time cosmic acceleration~\cite{70}. The most general diffeomorphism-invariant $f(R)$ action that one can write in 4D is given by~\cite{51,52}
\begin{equation}\label{70}
    S = M_p^2\int d^4x\sqrt{-g}\;f(R)
\end{equation}
where $M_p$ is the 4D Planck mass scale and 
\begin{equation}\label{71}
    f(R) = \sum_{i = -\infty}^{\infty}\beta_iR^i
\end{equation}
contains the Einstein-Hilbert term ($i=1$) along with both positive and inverse power corrections. The coefficients $\beta_i$ are of appropriate dimensions and scaling factors as suggested by effective field theory. There are various works in $f(R)$ theories over the years~\cite{53,54,55,56,57,58}. Now the action in eq.(\ref{70}) is dynamically equivalent to the following action with a new auxiliary field $\chi$
\begin{equation}\label{72}
    S = M_p^2\int d^4x \sqrt{-g}\left[f'(\chi)(R-\chi) + f(\chi)\right]
\end{equation}
One can verify that variation with respect to $\chi$ gives $f''(\chi)(R-\chi) = 0$ which implies that $\chi = R$ since $f''(\chi) \neq 0$ due to non-linear $f$ and thus one reproduces the starting action (\ref{70}). Now redefine the field $\chi$ by $\phi = f'(\chi)$ and define the potential of $\phi$ field as $V(\phi) = \chi(\phi)\phi - f(\chi(\phi))$. Then the action in (\ref{70}) becomes
\begin{equation}\label{73}
    S_J = M_p^2 \int d^4x\sqrt{-g}\left[\phi R - V(\phi)\right]
\end{equation}
This is the action expressed in the Jordan frame. Hence the $f(R)$ theory is dynamically equivalent to a Brans-Dicke theory with BD parameter $\omega = 0$ (as there is no kinetic piece in the action (\ref{73})). Now consider the conformal transformation of the form $g_{\mu\nu} \to e^{\sigma/\sqrt{3}M_p}g_{\mu\nu}$, where $e^{\sigma/\sqrt{3}M_p} \equiv \phi = f'(R)$. Under this conformal transformation, the action in (\ref{73}) can be written as
\begin{equation}\label{74}
    S_E = \int d^4x\sqrt{-g_E}\left[M_p^2R_E - \frac{1}{2}g_E^{\mu\nu}\partial_\mu\sigma\partial_\nu\sigma - U(\sigma)\right]
\end{equation}
which is the action in Einstein frame and $R_E$ and $g_E$ are the Einstein frame Ricci scalar and metric, respectively. The Einstein frame potential for the scalar field $\sigma$ is given as
\begin{equation}\label{75}
    U(\sigma) = M_p^2\frac{Rf'(R) - f(R)}{(f'(R))^2}
\end{equation}
Essentially, $f(R)$ theory has just one extra scalar degree of freedom with respect to GR. And this extra degree of freedom is actually a dynamical degree of freedom. Now consider the matter into the picture and add a matter action $S_M$ to the action in (\ref{74}). Then the full action is given by
\begin{equation}\label{76}
    S_E = \int d^4x\sqrt{-g_E}\left[M_p^2R_E - \frac{1}{2}g_E^{\mu\nu}\partial_\mu\sigma\partial_\nu\sigma - U(\sigma)\right] + S_M(e^{-\sigma/\sqrt{3}M_p}g^E_{\mu\nu}, \psi)
\end{equation}
where $\psi$ represents the matter fields. Recalling the analysis in the previous subsection. We identify the function $\mathcal A$ to be
\begin{equation}\label{77}
    \mathcal A^2(\sigma) = e^{-\sigma/\sqrt{3}M_p}
\end{equation}
Then the matter coupling function $\alpha(\sigma)$ is given by
\begin{equation}\label{78}
    \alpha(\sigma) = -\frac{1}{2\sqrt{3}M_p}
\end{equation}
See that the matter coupling function $\alpha(\sigma)$ is constant and thus its derivative with respect to $\sigma$ vanishes, that is, $\alpha'(\sigma) = 0$. Again we do the first order scalar perturbation around some $\sigma = \sigma_0$ with $U'(\sigma_0) = 0$ in order to find the effective mass squared of the scalar field $\sigma$. As $\alpha'(\sigma)$ vanishes everywhere, the effective mass squared is solely dependent on the second derivative of the Einstein frame scalar potential, that is the bare mass of the scalar field and is given by
\begin{equation}\label{79}
    \mu_{eff}^2 = U''(\sigma_0) = \frac{1}{f'(R_0)}\Bigg\{\frac{1}{3}\left[\frac{f'(R_0)}{f''(R_0)} - R_0\right]\Bigg\}
\end{equation}
where $f'(R_0) = e^{\sigma_0/\sqrt{3}M_p}$. The effective mass of the scalar field in eq.(\ref{79}) is also termed as the \textit{scalaron mass} in $f(R)$ theory and the term in the curly brackets is denoted by $m_\sigma^2$. Note that from action (\ref{73}), the effective Planck scale mass can be expressed as $M_{eff}^2 = M_p^2f'(R)$, hence $f'(R)$ should always be positive in order to avoid the ghost instability. Thus, the sign of $m_\sigma^2$ determine the tachyonic (in)stability of the scalaron. An important point to notice is that the tachyonic (in)stability of the scalaron does not depend on the matter content and also which type of matter it is. The (in)stability depends purely on the construction of $f(R)$ theory. For a simple example, consider the Starobinsky $f(R)$ model, which incorporates inflation in the early universe without introducing an inflaton field by hand~\cite{59,60} : $f(R) = R + \beta R^2$, where $\beta$ is a parameter of the model with inverse square mass dimensions. One can easily verify the scalaron mass of Starobinsky model is $m_\sigma^2 = 1/6\beta$. Hence in order to avoid the tachyonic instability of the scalaron, one needs to impose the following constraint on the parameter $\beta > 0$. Generally the spontaneous scalarization occurs in compact objects through the matter-induced tachyonic instability and exhibits a threshold behavior. But in $f(R)$ gravity, the scalaron, the extra scalar degree of freedom can be determined by the chosen $f(R)$ model, although its dynamics are still influenced by the matter background through the trace of the energy-momentum tensor. Therefore, the notion of spontaneous scalarization should be used with caution in $f(R)$ theories.\\

\noindent The next section \ref{RS section} contains the main calculations and results of this work. We will show the 5D warped braneworld gravity can be written as an effective on-brane scalar-tensor theory with the extra-dimensional modulus field playing the role of the scalar degree of freedom and extensively use the machinery developed for scalar-tensor theory in section \ref{ST} in order to discuss the tachyonic instability and scalarization in braneworld setup.

\section{(In)stability in Randall-Sundrum Braneworlds}\label{RS section}

\subsection{Effective 4D Action on RS Branes}\label{effective}

We briefly review the background metric discovered by Randall and Sundrum to set up our notations. The whole set up is a 5D Einstein gravity with a bulk cosmological constant $\Lambda_b$ with two 3-branes located in the orbifold $S^1/\mathbb Z_2$ at the fifth dimensional coordinate $z = 0$ (Planck brane or hidden brane) and $z = r_c$ (TeV brane or visible brane). The full 5D gravitational action~\cite{3}
\begin{equation}\label{17}
S = 2\int d^4x\int_0^{r_c}dz\sqrt{-g_5}\;[M_*^3R_5 - \Lambda_b] - T_{hid}\int d^4x \sqrt{-g_{(+)}} - T_{vis}\int d^4x \sqrt{-g_{(-)}}
\end{equation}
where, $M_*$ is the 5D Planck mass scale defined in term of 5D graviational constant $G_5$ as $M_*^{-3} = 8\pi G_5$. $T_{hid,vis}$ are the brane tensions of hidden brane and of visible brane. The induced metric on positive tension brane is $g_{(+)\mu\nu} = g_{\mu\nu}$ and on negative tension brane is $g_{(-)\mu\nu} = e^{-2kr_cT}$. Note the range of the extra dimension in $S^1/\mathbb Z_2$ is $z \in [-r_c,r_c]$ with $z$ and $-z$ identified. Randall and Sundrum showed that there exists a metric solution which respects the 4D Poincar\'e invariance
\begin{equation}\label{18}
    ds^2 = e^{-2k|z|}\eta_{\mu\nu}dx^\mu dx^\nu + dz^2
\end{equation}
if and only if the following \textit{fine-tuning conditions} are satisfied
\begin{equation}\label{19}
    T_{hid} = -T_{vis} = - \frac{\Lambda_b}{2k}\;\;\;\;;\;\;\;\; \Lambda_b = -12M_*^3k^2
\end{equation}
To include the massless gravitational degrees of freedom (the zero modes about the background spacetime in (\ref{18})), we replace Minkowski metric $\eta_{\mu\nu}$ with a general metric $g_{\mu\nu}$ and the $z$-direction length $r_c$ with a modulus or radion field $T(x)$
\begin{equation}\label{20}
    ds^2 = e^{-2kT(x)|z|}g_{\mu\nu}(x)dx^\mu dx^\nu + T^2(x)dz^2
\end{equation}
Since we impose $\mathbb Z_2$ symmetry ($z \to -z$), massless vector fluctuations associated with the off-diagonal part of the metric are absent. \\
The effective 4D theory on either brane is of Brans-Dicke (BD) form with the following action~\cite{22,26}
\begin{equation}\label{21}
    S = \int d^4x \sqrt{-g_{(\pm)}}\frac{1}{16\pi} \left(\Phi_{\pm}R - \frac{\omega_\pm(\Phi_\pm)}{\Phi_\pm}\nabla_\mu\Phi_\pm\nabla^\mu\Phi_\pm\right)
\end{equation}
The Brans-Dicke scalar field (or the inverse of the effective gravitational constant) is
\begin{equation}\label{22}
    \frac{1}{G^\pm_{eff}} = \Phi_{\pm} = \frac{32\pi M^3}{k}e^{\mp kr_cT}\sinh(kr_cT)
\end{equation}
while the corresponding Brans-Dicke function is
\begin{equation}\label{23}
    \omega_{\pm}(T) = \pm 3e^{\pm kr_cT}\sinh(kr_cT)
\end{equation}
where we denote the Planck brane (TeV brane) BD scalar as $\Phi_+$ ($\Phi_-$) and the ranges of BD functions are $0 < \omega_+ < \infty$ for the Planck Brane and $-3/2 < \omega_- < 0$ for the TeV Brane. In terms of the BD scalar, the BD functions can be written as
\begin{equation}\label{24}
    \omega_+(\Phi_+) = \frac{3}{2}\frac{\Phi_+}{\gamma - \Phi_+}\;\;\;;\;\;\; \omega_-(\Phi_-) = -\frac{3}{2}\frac{\Phi_-}{\gamma + \Phi_-}
\end{equation}
where $\gamma = 16\pi M^3/k$. The derivatives with respect to corresponding BD scalars are
\begin{equation}\label{25}
    \omega'_+(\Phi_+) = \frac{3}{2}\frac{\gamma}{(\gamma - \Phi_+)^2}\;\;\;;\;\;\; \omega'_-(\Phi_+) = - \frac{3}{2}\frac{\gamma}{(\gamma + \Phi_-)^2}.
\end{equation}
Using eq.(\ref{23}) and (\ref{24}) we obtain
\begin{equation}\label{26}
    e^{2kr_cT} = \frac{\gamma}{\gamma - \Phi_+} = 1 + \frac{\Phi_-}{\gamma}
\end{equation}
which imply the ranges of $\Phi_\pm$ are
\begin{equation}\label{27}
    0 \leq \Phi_+ < \gamma \;\;\;;\;\;\; 0 \leq \Phi_- < \infty
\end{equation}
Here the action in eq.(\ref{21}) is expressed in Jordan frame. We want to present our analysis same as section \ref{ST} in Einstein frame. Conformally transform the metric as follows
\begin{equation}\label{28}
    g_{\mu\nu} \to \tilde g_{\mu\nu} = \Phi_{\pm}g_{\mu\nu}
\end{equation}
But as mentioned earlier, one can equivalently work in Jordan frame to do all the analysis. Here $\sqrt{-\tilde g} = \Phi_\pm^2\sqrt{-g}$. Also the conformally transformed Ricci scalar is given by
\begin{equation}\label{29}
    R = \Phi_\pm \left(\tilde R - 3\tilde\Box(\ln\Phi_\pm) - \frac{3}{2}\tilde\nabla_\mu(\ln\Phi_\pm)\tilde\nabla^\mu(\ln\Phi_\pm)\right)
\end{equation}
where the $\tilde R$ is the Einstein frame Ricci scalar and $\tilde\Box,\tilde\nabla$ are with respect to the Einstein frame metric $\tilde g$. In eq.(\ref{29}), the second term becomes a total derivative. Therefore modulo the surface contributions, the Einstein frame action is given by 
\begin{equation}\label{30}
    S_E = \int d^4x \sqrt{-\tilde g} \frac{1}{16\pi}\left[\tilde R - \frac{3 + 2\omega_\pm(\Phi_\pm)}{2}\tilde\nabla_\mu(\ln\Phi_\pm)\tilde\nabla^\mu(\ln\Phi_\pm)\right]
\end{equation}
We redefine the scalar field in order to make the kinetic term canonical
\begin{equation}\label{31}
    \frac{d\varphi_\pm}{d\ln\Phi_\pm} = \sqrt{\frac{3 + 2\omega_\pm(\Phi_\pm)}{16\pi}}
\end{equation}
Then the Einstein frame action in terms of redefined scalar field $\varphi_\pm$ is given by
\begin{equation}\label{32}
    S_E = \int d^4x \sqrt{-\tilde g} \frac{1}{16\pi}\left[\tilde R - \frac{1}{2}\tilde\nabla_\mu\varphi_\pm\tilde\nabla^\mu\varphi_\pm\right]
\end{equation}
Up to now we have not considered any matter contribution to the action. Now we consider localized matter in both the branes (i.e., both Planck and TeV brane). We denote the Planck brane (TeV brane) matter action as $S_M^{(+)}$ ($S_M^{(-)}$).
\begin{equation}\label{33}
    S_E = \int d^4x \sqrt{-\tilde g}\frac{1}{16\pi}\left[\tilde R - \frac{1}{2}\tilde\nabla_\mu\varphi_\pm\tilde\nabla^\mu\varphi_\pm\right] + S^{(\pm)}_M(\Phi_\pm^{-1}\tilde g_{\mu\nu},\psi)
\end{equation}
From the knowledge of previous section \ref{ST}, we identify the function $\mathcal A$ as
\begin{equation}\label{34}
    \mathcal A^2(\Phi_\pm) = \Phi_\pm^{-1}
\end{equation}
One can easily notice that in Jordan frame action (\ref{21}) or in Einstein frame action (\ref{33}), there is no potential term corresponding to $\Phi_\pm$ or $\varphi_\pm$. This is because the fine-tuning conditions in (\ref{19}) identically make the potential term to vanish. If the fine-tunings are slightly detuned, one can generate a potential term for $\Phi_\pm$ or $\varphi_\pm$ purely from gravitational sector and we denote that potential term in Jordan frame as $V_{GR}$. Also one can generate a potential term by introducing a bulk scalar field which stabilizes the radion field (according to Goldberger-Wise mechanism of radion stabilization) and we denote the potential term coming from bulk stabilizing field as $V_{GW}$ in Jordan frame. Therefore we can generalize the action in (\ref{33}) for the three cases (tuned-RS, detuned-RS and GW stabilized RS) as
\begin{equation}\label{35}
    S_E = \int d^4x \sqrt{-\tilde g}\frac{1}{16\pi}\left[\tilde R - \frac{1}{2}\tilde\nabla_\mu\varphi_\pm\tilde\nabla^\mu\varphi_\pm - U_i(\varphi_\pm)\right] + S^{(\pm)}_M(\Phi_\pm^{-1}\tilde g_{\mu\nu},\psi)
\end{equation}
where $U_i(\varphi_\pm) = 16\pi V_i(\Phi_\pm)/\Phi_\pm^2$ is the Einstein frame potential with
\[
V_i(\Phi_\pm) =
\begin{cases}
0 & ;\;\;\text{Tuned-RS} \\
V_{GR}(\Phi_\pm)  & ;\;\;\text{Detuned-RS} \\
V_{GW}(\Phi_\pm)  & ;\;\;\text{GW Stabilized-RS}
\end{cases}
\]
Using the eq.(\ref{11}), (\ref{31}) and (\ref{34}) we get the following matter coupling function in terms of Einstein frame radion field $\varphi_\pm$
\begin{equation}\label{36}
    \alpha(\varphi_\pm) = - \sqrt{\frac{4\pi}{3 + 2\omega_\pm(\Phi_\pm)}}
\end{equation}

\noindent The equation of motion of the Einstein frame radion field $\varphi_\pm$ is given by
\begin{equation}
    \Box\varphi_\pm = U_{eff}^{(i)'}(\varphi_\pm)
\end{equation}
where, $U_{eff}^{(i)}(\varphi_\pm)$ is the effective Einstein frame radion potential, which is given by
\begin{equation}
    U_{eff}^{(i)}(\varphi_\pm) = U_i(\varphi_\pm) - 16\pi T_{(\pm)}\int d\varphi_\pm\;\alpha(\varphi_\pm)
\end{equation}
where $T_{(+)}\; (T_{(-)})$ denotes the trace of Einstein frame energy-momentum tensor of matter in Planck brane (TeV brane). One can perform the above integration of $\alpha(\varphi_\pm)$ with respect to $\varphi_\pm$ using the eq.(\ref{24}), (\ref{31}) and (\ref{36}) and obtain the following result for both the Planck ($+$) and TeV branes ($-$)~\footnote{see Appendix \ref{A} for the derivation}
\begin{equation}\label{a39}
    \int d\varphi_+\;\alpha(\varphi_+) = -\frac{1}{2}\ln\left[\gamma\;\text{sech}^2\left(\sqrt{\frac{4\pi}{3}}\varphi_+\right)\right] \;\;\;\; ; \;\;\;\; \int d\varphi_-\;\alpha(\varphi_-) = -\frac{1}{2}\ln\left[\gamma\;\text{csch}^2\left(\sqrt{\frac{4\pi}{3}}\varphi_-\right)\right]
\end{equation}
where, we have ignored the integration constants, as these will only add a constant shift to the potential, hence can be ignored safely. Now the effective mass squared of the Einstein frame radion field at some $\varphi_\pm = \varphi_{\pm 0}$ is given by
\begin{equation}
    \mu_{eff}^2 \equiv U_{eff}^{(i)''}(\varphi_{\pm 0})
\end{equation}
if in addition to it, $U_{eff}^{(i)'}(\varphi_{\pm 0}) = 0$ has physical solutions, then $\varphi_\pm = \varphi_{\pm 0}$ will be one of the extremum of $U_{eff}^{(i)}$ (can be minimum, maximum or inflection point).

\subsubsection*{General Remarks on Radion Perturbations and Tachyonic Instability}

The Einstein frame Radion field $\varphi_\pm$ is a function of on-brane spacetime points, that is, $\varphi_\pm = \varphi_\pm(x^\mu)$. Hence the scalar perturbation $\delta\varphi_\pm(x^\mu)$ is also a function of on-brane spacetime, in general. For any general curved spacetime, the decomposition of the radion fluctuations in the eigenmode basis entirely depends on the on-brane background geometry through the $\Box = g^{\mu\nu}\nabla_\mu\nabla_\nu$ operator corresponding to that geometry. For example, in flat spacetime the eigenbasis is plane waves and in spherically symmetric spacetimes the eigenbasis is spherical harmonics and so on. In the present work, however, our objective is not to determine the complete spectrum of unstable modes, but rather to identify the onset of the instability through the sign of the effective mass squared. Although for clarity we provide below a brief discussion about the unstable radion modes in flat on-brane geometry.\\

\noindent The radion fluction satisfies the following equation in general (we suppress the $\pm$ subscripts for simplicity)
\begin{equation}
    \left(\Box - \mu_{eff}^2\right)\delta\varphi = 0
\end{equation}
In flat spacetime, we can decompose $\delta\varphi$ in plane wave modes
\begin{equation}
    \delta\varphi(t,\mathbf x) = \int \frac{d^3k}{(2\pi)^3}e^{i\mathbf k\cdot\mathbf x}\delta\varphi_k(t) 
\end{equation}
which yields the time-evolution equation for each momentum modes of the radion field
\begin{equation}
    \delta\ddot\varphi_k(t) + (k^2+\mu^2_{eff})\delta\varphi_k(t) = 0
\end{equation}
Hence, for $\mu_{eff}^2<0$, the low-momentum modes $k^2<|\mu_{eff}^2|$ grow exponentially signaling a tachyonic instability of the mode. In general curved spacetimes, the plane wave basis is replaced by the appropriate eigenfunctions of the background geometry. But the onset of the instability is still signaled by the appearance of a negative effective mass squared, which provides the necessary condition for a tachyonic instability. The precise unstable spectrum, however, depends on the background geometry and the associated mode equation. When referring to a spatially homogeneous radion field, we mean the lowest spatial mode of the corresponding background geometry, which reduces to the zero-momentum $k=0$ mode in flat spacetime. In section \ref{flrw}, we discussed the tachyonic instability of the zero-mode of the radion field in positively curved (dS) and negatively curved (AdS) backgrounds. In the subsequent sections, our analysis of the Tuned-RS, Detuned-RS and GW-stabilized RS scenarios is performed for a spatially homogeneous radion background. The (in)stability criterion is therefore derived for this homogeneous configuration, while the discussion above clarifies how perturbations about this background should be interpreted.

\subsection{Tuned RS}\label{Tuned-RS}

In this section, we will study the tachyonic (in)stability of radion field in Einstein frame with the fine-tuning conditions in eq.(\ref{19}).
The derivative of $\alpha(\varphi_\pm)$ with respect to $\varphi_\pm$ is given by
\begin{equation}\label{37}
    \alpha'(\varphi_\pm) = \frac{8\pi\Phi_{\pm}\omega_\pm'(\Phi_\pm)}{(3 + 2\omega_\pm(\Phi_\pm))^2}
\end{equation}
where, $\omega'_\pm$ is the derivative of $\omega_\pm$ with respect to $\Phi_\pm$. From eq.(\ref{27}) we know $\Phi_\pm \geq 0$ always. Therefore the sign of $\alpha'(\varphi_\pm)$ is determined through the sign of $\omega'_\pm(\Phi_\pm)$. Now from eq.(\ref{25}) we can see that for Planck brane (TeV brane) $\omega'_+ > 0$ ($\omega_-'< 0$). Using the similar analysis in section \ref{ST}, we can get the effective mass squared for the Einstein frame radion field (doing first order perturbation around some $\varphi_\pm = \varphi_{\pm0}$ with $\alpha(\varphi_{0\pm}) = 0$)
\begin{equation}\label{38}
    \mu^2_{eff} = -16\pi\alpha'(\varphi_{0\pm})T_{(\pm)}
\end{equation}
where $T_{(+)}\; (T_{(-)})$ denotes the trace of energy-momentum tensor of matter in Planck brane (TeV brane). Below we discuss about the tachyonic (in)stability of radion field in the two branes in the context of tuned-RS considering different cases.

\subsubsection*{On Planck Brane}

For Planck brane, $\omega_+' > 0$ and hence $\alpha' > 0$. For vacuum or electrovacuum (or any conformal matter) $T_{(+)} = 0$ which implies $\mu_{eff}^2 = 0$. This suggests no tachyonic instability of radion field in Planck brane with no matter or conformal matter. Now for any non-relativistic matter or fluid we have $T_{(+)} \simeq -\rho < 0$, where $\rho$ is the energy density. For such matter in Planck brane, $\mu_{eff}^2 > 0$ suggests no tachyonic instability of radion field in Planck brane. Finally, if for some matter $T_{(+)} > 0$, then $\mu_{eff}^2 < 0$, then the radion field is prone to tachyonic instability in Planck brane. Although, the exact threshold for the tachyonic instability can only be determined if we specify any background geometry on Planck brane.\\

\begin{figure}[h!]
    \centering
    
    \begin{subfigure}[b]{0.32\textwidth}
        \centering
        \includegraphics[width=\textwidth]{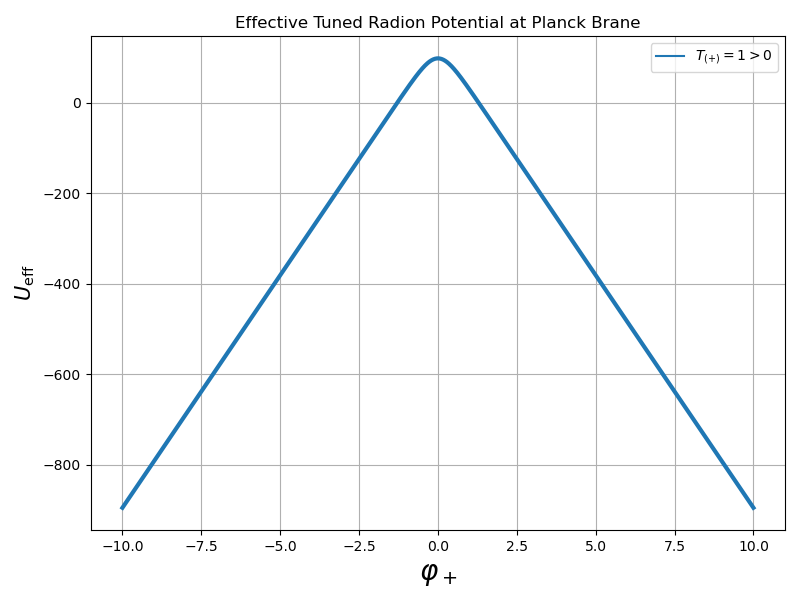}
        \caption{}
    \end{subfigure}
    \hfill
    \begin{subfigure}[b]{0.32\textwidth}
        \centering
        \includegraphics[width=\textwidth]{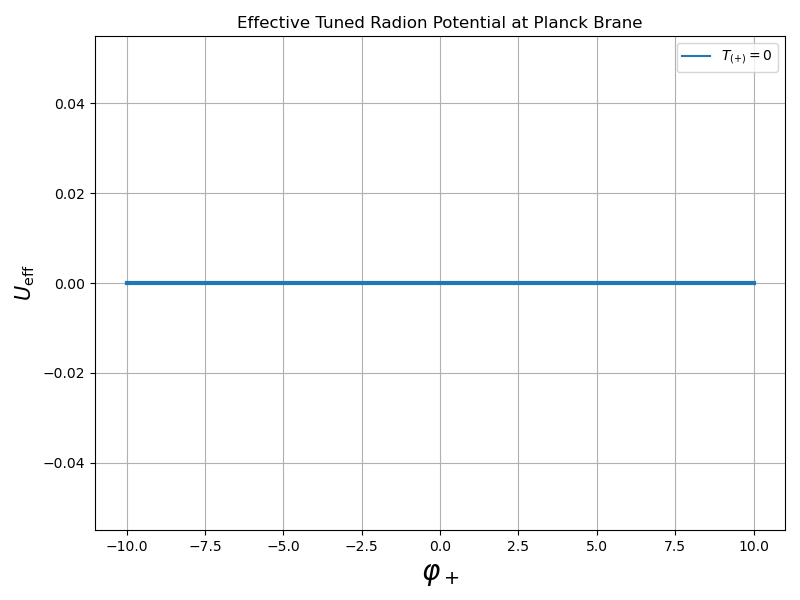}
        \caption{}
    \end{subfigure}
    \hfill
    \begin{subfigure}[b]{0.32\textwidth}
        \centering
        \includegraphics[width=\textwidth]{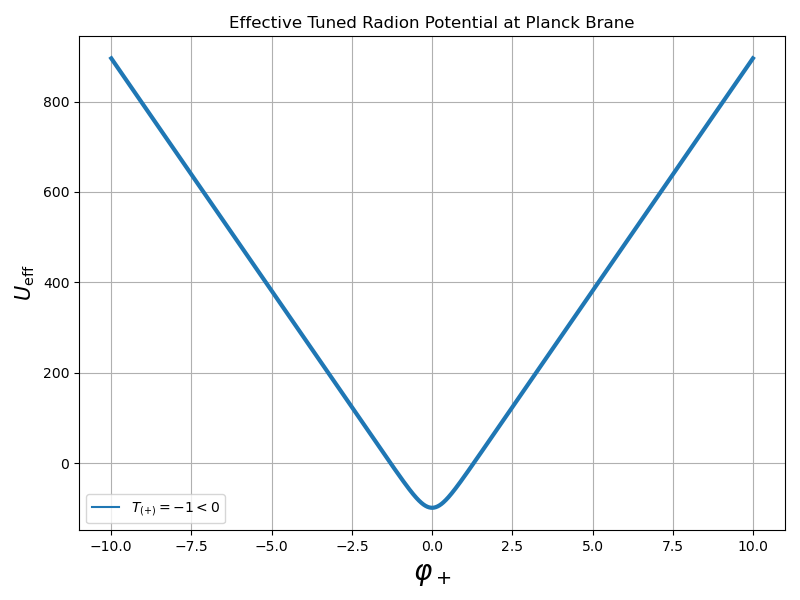}
        \caption{}
    \end{subfigure}
    
    \caption{Plots of the effective tuned radion potential $U_{eff}$ at Planck brane with respect to the Einstein frame radion field $\varphi_+$ for three cases $T_{(+)} >0,=0,<0$ respectively in (a), (b), (c). We have chosen $\gamma = 16\pi$ and $k = 1$ for obtaining these plots.}
    \label{fig:tuned_planck}
\end{figure}

\noindent One can see the above conclusions of tachyonic instability of radion field on Planck brane by explicitly writing the effective potential of $\varphi_+$ as
\begin{equation}
    U_{eff} = 8\pi T_{(+)}\ln\left[\gamma\;\text{sech}^2\left(\sqrt{\frac{4\pi}{3}}\varphi_+\right)\right]
\end{equation}
The plot of this effective potential with respect to $\varphi_+$ is given in Figure (\ref{fig:tuned_planck}).
We can see from the Figure that for $T_{(+)} = 0$, the effective potential vanishes. For both $T_{(+)}\neq0$ cases, $\varphi_+ = 0$ appears as a stationary point.\\

\noindent The effective mass squared as a function of $\varphi_+$ is expressed by the second derivative of $U_{eff}$
\begin{equation}
    \mu_{eff}^2 \equiv U''_{eff} = -\frac{64\pi^2T_{(+)}}{9}\text{sech}^2\left(\sqrt{\frac{4\pi}{3}}\varphi_+\right)
\end{equation}
Note that for Planck brane $U_{eff}' = 0$ does have a physical solution
\begin{equation}
    U'_{eff}(\varphi_+) = -16\pi T_{(+)}\sqrt{\frac{4\pi}{3}}\tanh\left(\sqrt{\frac{4\pi}{3}}\varphi_+\right) = 0 \implies \varphi_+ = 0
\end{equation}
So $\varphi_+ = 0$ is an extremum point and $\mu_{eff}^2 (\varphi_+=0) = U''_{eff}(\varphi_+=0) = - 64\pi^2 T_{(+)}/9$. For $T_{(+)} > 0$, $\mu_{eff}^2 < 0$, means it is a local maximum and for $T_{(+)} < 0$, $\mu_{eff}^2 > 0$, means it is a local minimum. While $T_{(+)} = 0$, $\mu_{eff}^2 = 0$, suggests that potential is flat here.

\subsubsection*{On TeV Brane}

For TeV brane, $\omega_-' < 0$ and hence $\alpha' < 0$. For vacuum or electrovacuum (or any conformal matter) $T_{(-)} = 0$ which implies $\mu_{eff}^2 = 0$. This suggests no tachyonic instability of radion field in TeV brane with no matter or conformal matter. Now for any non-relativistic matter or fluid we have $T_{(-)} \simeq -\rho < 0$, where $\rho$ is the energy density. For such matter in TeV brane, $\mu_{eff}^2 < 0$ suggests that the radion field in TeV brane is prone to tachyonic instability. Although, the exact threshold for the tachyonic instability can only be determined if we specify any background geometry on TeV brane. Finally, if for some matter $T_{(-)} > 0$, then $\mu_{eff}^2 > 0$, then the radion field has no tachyonic instability in TeV brane.\\

\begin{figure}[h!]
    \centering
    
    \begin{subfigure}[b]{0.32\textwidth}
        \centering
        \includegraphics[width=\textwidth]{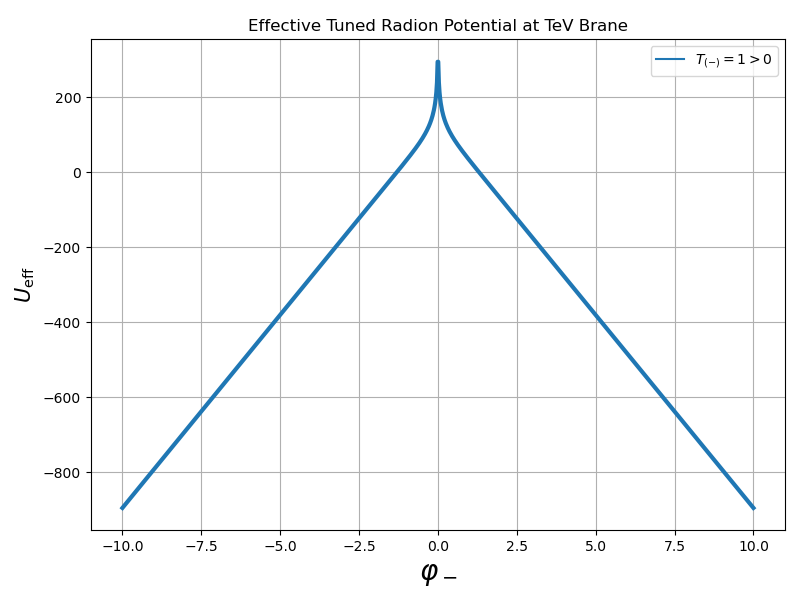}
        \caption{}
    \end{subfigure}
    \hfill
    \begin{subfigure}[b]{0.32\textwidth}
        \centering
        \includegraphics[width=\textwidth]{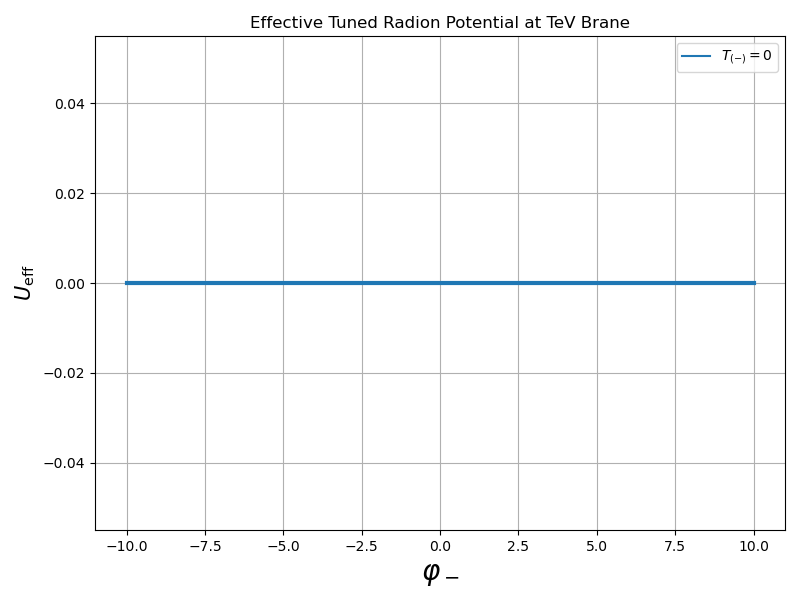}
        \caption{}
    \end{subfigure}
    \hfill
    \begin{subfigure}[b]{0.32\textwidth}
        \centering
        \includegraphics[width=\textwidth]{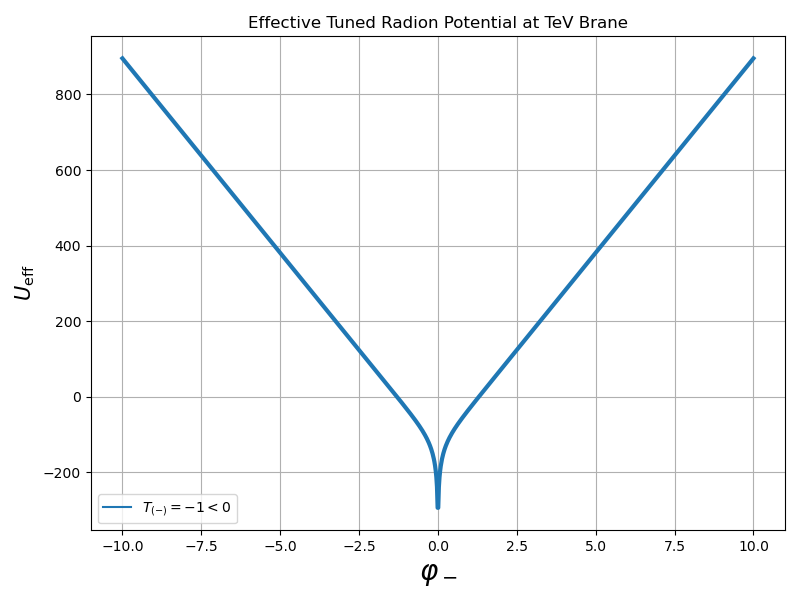}
        \caption{}
    \end{subfigure}
    
    \caption{Plots of the effective tuned radion potential $U_{eff}$ at TeV brane with respect to the Einstein frame radion field $\varphi_-$ for three cases $T_{(-)} >0,=0,<0$ respectively in (a), (b), (c). We have chosen $\gamma = 16\pi$ and $k = 1$ for obtaining these plots.}
    \label{fig:tuned_tev}
\end{figure}

\noindent Similar to the Planck brane analysis, one can also see the above conclusions of tachyonic instability of radion field on TeV brane by explicitly writing the effective potential of $\varphi_-$ as
\begin{equation}
    U_{eff} = 8\pi T_{(-)}\ln\left[\gamma\;\text{csch}^2\left(\sqrt{\frac{4\pi}{3}}\varphi_-\right)\right]
\end{equation}
The plot of this effective potential with respect to $\varphi_-$ is given in Figure (\ref{fig:tuned_tev}).
We can see from the Figure that for $T_{(-)} = 0$, the effective potential vanishes. For both $T_{(-)}\neq0$ cases, there is a logarithmic singularity at $\varphi_- = 0$ due to the presence of $\ln\text{csch}^2$ term in the effective potential.\\

\noindent The effective mass squared as a function of $\varphi_-$ is expressed by the second derivative of $U_{eff}$
\begin{equation}
    \mu_{eff}^2 \equiv U_{eff}'' = \frac{64\pi^2T_{(-)}}{9}\text{csch}^2\left(\sqrt{\frac{4\pi}{3}}\varphi_-\right)
\end{equation}
Note that for TeV brane $U_{eff}' = 0$ does not have a physical solution, hence there are no extremum points of $U_{eff}$ at the TeV brane.\\

\noindent One of the criteria for spontaneous scalarization to occur is that there should be a tachyonic instability in first order scalar perturbation. And spontaneous scalarization generally occur in compact objects that means $T \simeq -\rho < 0$, where $\rho$ is the energy density. Therefore the scalarization process will be more feasible in TeV brane rather than Planck brane with the fine-tuning conditions. In the next subsection, we will discuss the tachyonic instability of radion field on either brane for the case of detuning of RS parameters.

\subsection{Detuned RS}\label{detuned-RS}

In getting the action in eq.(\ref{21}) as the 4D effective theory, we assumed the fine-tuning conditions (\ref{19}). This made the gravitational contribution to radion potential vanish. However, we could take the brane tensions to be detuned and then the non-vanishing radion potential would alter the tachyonic (in)stability criteria of the on-brane scalar-tensor theory. Defining $\phi \equiv Ae^{-kr_cT(x)}$, where $A^2 = 12M_*^3/k$, to be the canonical radion field. Finally one obtains the following form for the gravitational contribution to radion potential on the Planck brane
\begin{equation}\label{39}
    V_{GR}(\phi) = \frac{k^4}{A^4}\left(\tau\phi^4 + \Lambda_{4D}\right)
\end{equation}
where $\tau = \left(T_{vis} - \frac{\Lambda_b}{2k}\right)/k^4$ and $\Lambda_{4D}$ is the 4D cosmological constant on the Planck brane given as
\begin{equation}\label{40}
    \Lambda_{4D} = \frac{A^4}{k^4}\left(T_{hid}+\frac{\Lambda_b}{2k}\right)
\end{equation}
Now the TeV brane action is obtained from the Planck brane action by the conformal transformation $g_{\mu\nu(-)} = e^{-2kr_cT}g_{\mu\nu(+)}$. Hence the TeV brane radion potential is obtained by multiplying eq.(\ref{39}) with $e^{4kr_cT}$. Thus the TeV brane gravitational radion potential is
\begin{equation}\label{41}
    V_{GR}(\phi) = \frac{k^4}{A^4}\left(\Lambda_{4D}' + \frac{\tau'}{\phi^4}A^4\right)
\end{equation}
where $\tau' = \Lambda_{4D}$ and $\Lambda_{4D}' = \tau A^4$ is the TeV brane cosmological constant. For Planck brane $\tau,\Lambda_{4D}$ and for TeV brane $\tau',\Lambda_{4D}'$ are called the \textit{detuning parameters} of RS model. And $\tau = \Lambda_{4D} = 0$ or $\tau' = \Lambda_{4D}' = 0$ gives back the fine-tuning conditions in eq.(\ref{19}). Below we discuss the consequences of the radion potential to the tachyonic instability of radion field on the either branes.

\subsubsection*{On Planck Brane}

Using eq.(\ref{22}) we can write the radion potential in term of the BD scalar $\Phi_+$ on the Planck brane as
\begin{equation}\label{42}
    V_{GR}(\Phi_+) = \frac{k^4}{A^4}\left[\tau\left(1 - \frac{\Phi_+}{\gamma}\right)^2A^4 + \Lambda_{4D}\right]
\end{equation}
The corresponding Einstein frame potential is given by
\begin{equation}\label{43}
    U_{GR}(\varphi_+) \equiv \frac{16\pi}{\Phi_+^2}V_{GR}(\Phi_+)
\end{equation}
We can express the BD scalar $\Phi_+$ in terms of the Einstein frame radion field $\varphi_+$ as~\footnote{see Appendix \ref{A} for the derivation}
\begin{equation}\label{a51}
    \Phi_+ = \gamma\;\text{sech}^2\left(\sqrt{\frac{4\pi}{3}}\varphi_+\right)
\end{equation}
Then the complete effective potential of radion field at the Planck brane in terms of Einstein frame radion field $\varphi_+$ is given by
\begin{equation}\label{u_p}
    U_{eff}^{(GR)}(\varphi_+) = \tau_*\sinh^4\left(\sqrt{\frac{4\pi}{3}}\varphi_+\right) + \Lambda_*\cosh^4\left(\sqrt{\frac{4\pi}{3}}\varphi_+\right) + T^+_*\ln\text{sech}^2\left(\sqrt{\frac{4\pi}{3}}\varphi_+\right) - T_*^+\ln\gamma
\end{equation}
where, $\tau_* \equiv 16\pi k^4\tau/\gamma^2,\;\; \Lambda_* \equiv 16\pi k^4\Lambda_{4D}/A^4\gamma^2,\;\;T_*^+ \equiv 8\pi T_{(+)}$.
Again by doing the similar analysis as in section \ref{ST}, that is, linearize the Einstein frame radion field at first order around some $\varphi_+$, we get the effective mass squared for the first order scalar perturbation $\mu_{eff}^2 = U^{(GR)''}_{eff}$. Assuming $z = \sqrt{\frac{4\pi}{3}}\varphi_+$ and $s \equiv \sinh^2z \geq 0$, we write the effective potential in terms of $s$
\begin{equation}
    U_{eff}^{(GR)} = (\tau_*+\Lambda_*)s^2 + 2\Lambda_*s + \Lambda_* - T_*^+\ln(1+s) - T_*^+\ln\gamma
\end{equation}
Here, $U_{eff}^{(GR)'} = 0$ does have solutions. The $U_{eff}^{(GR)'}$ is given as
\begin{align}
U_{eff}^{(GR)'}
&= \sqrt{\frac{4\pi}{3}}2\tanh z\left[2(\tau_*+\Lambda_*)s^2 + 2(\tau_*+2\Lambda_*)s + (2\Lambda_*-T_*^+)\right]
\end{align}
Clearly, $U_{eff}^{(GR)'} = 0$ solutions are
\begin{equation}\label{phi_0}
    z = 0 \implies \varphi_+ = 0
\end{equation}
and
\begin{equation}
    s_\pm = \frac{-2(\tau_*+2\Lambda_*) \pm \sqrt{\tau_*^2+2T_*^+(\tau_*+\Lambda_*)}}{2(\tau_*+\Lambda_*)} \implies \varphi_+ = \pm \sqrt{\frac{3}{4\pi}}\sinh^{-1}\sqrt{s_\pm}
\end{equation}
The last solution is valid only when $s_\pm \geq 0$, $\tau_*+\Lambda_* \neq 0$ and $\tau_*^2+2T_*^+(\tau_*+\Lambda_*) \geq 0$. The effective mass squared of the Einstein frame radion field at the Planck brane on these extremum points can be given by
\begin{equation}
    \mu_{eff}^2\Bigg|_{\varphi_+ = 0} = \frac{8\pi}{3}(2\Lambda_* - T_*^+)
\end{equation}
Thus, for $2\Lambda_* > T_*^+$, $\mu_{eff}^2 > 0$, means no tachyonic instability of the radion field (a local minimum) and for $2\Lambda_* < T_*^+$, $\mu_{eff}^2 < 0$, means radion field is prone to tachyonic instability (a local maximum) at $\varphi_+ = 0$. While for $2\Lambda_* = T_*^+$, $\mu_{eff}^2 = 0$, means the effective potential is flat at $\varphi_+ = 0$ (no tachyonic instability).\\

\begin{figure}[h!]
    \centering
    
    \begin{subfigure}[b]{0.32\textwidth}
        \centering
        \includegraphics[width=\textwidth]{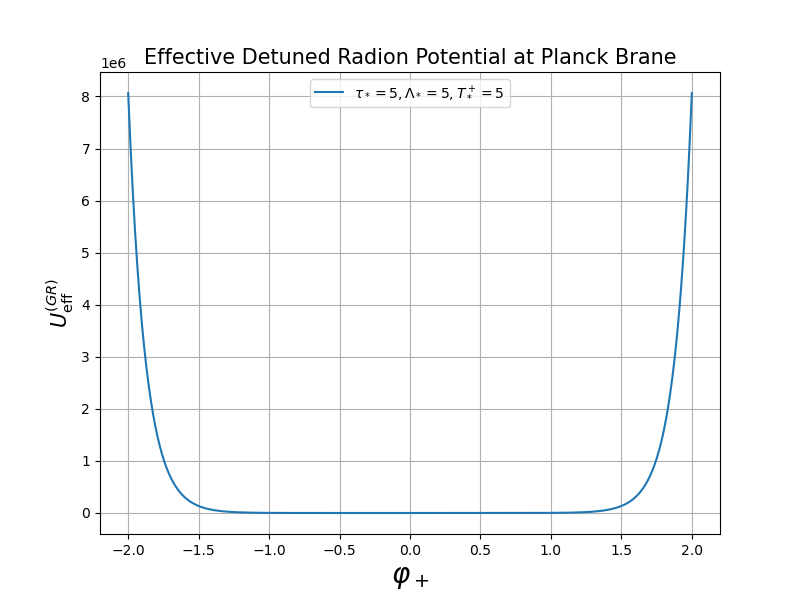}
        \caption{}
    \end{subfigure}
    \hfill
    \begin{subfigure}[b]{0.32\textwidth}
        \centering
        \includegraphics[width=\textwidth]{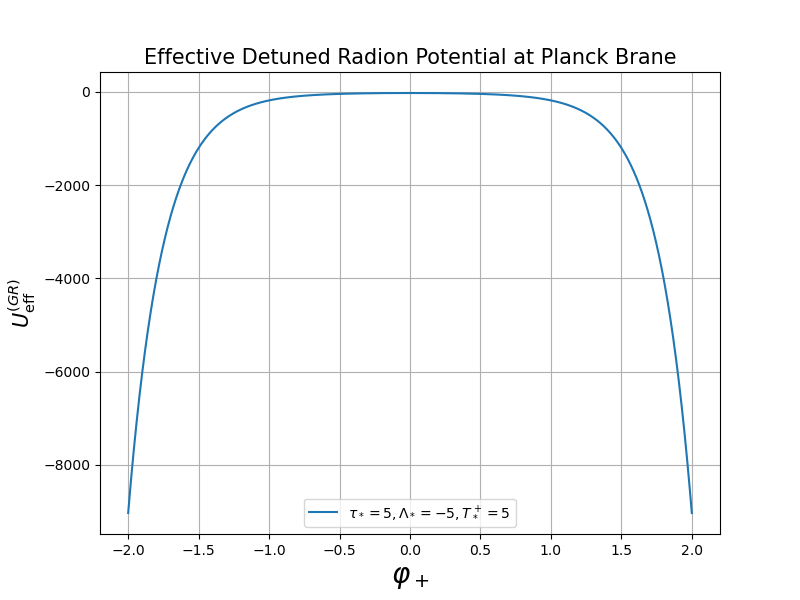}
        \caption{}
    \end{subfigure}
    \hfill
    \begin{subfigure}[b]{0.32\textwidth}
        \centering
        \includegraphics[width=\textwidth]{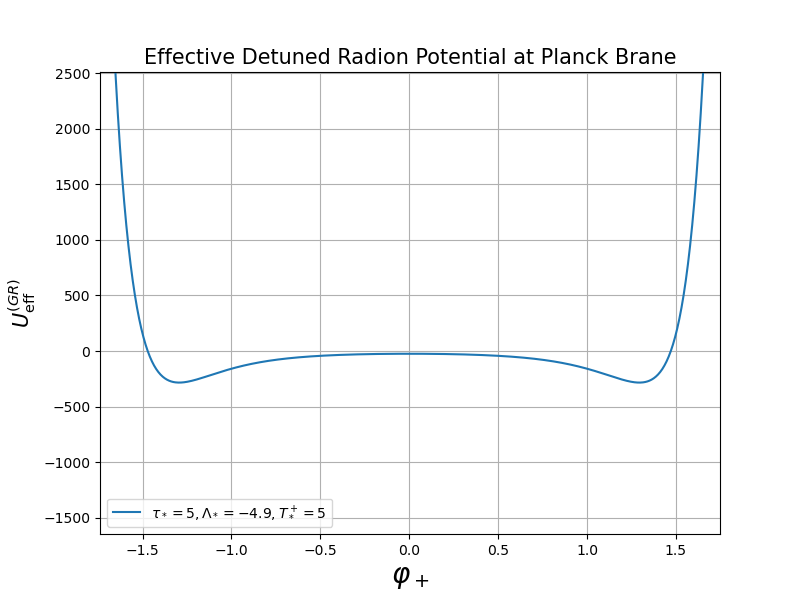}
        \caption{}
    \end{subfigure}
    \\
    \begin{subfigure}[b]{0.32\textwidth}
        \centering
        \includegraphics[width=\textwidth]{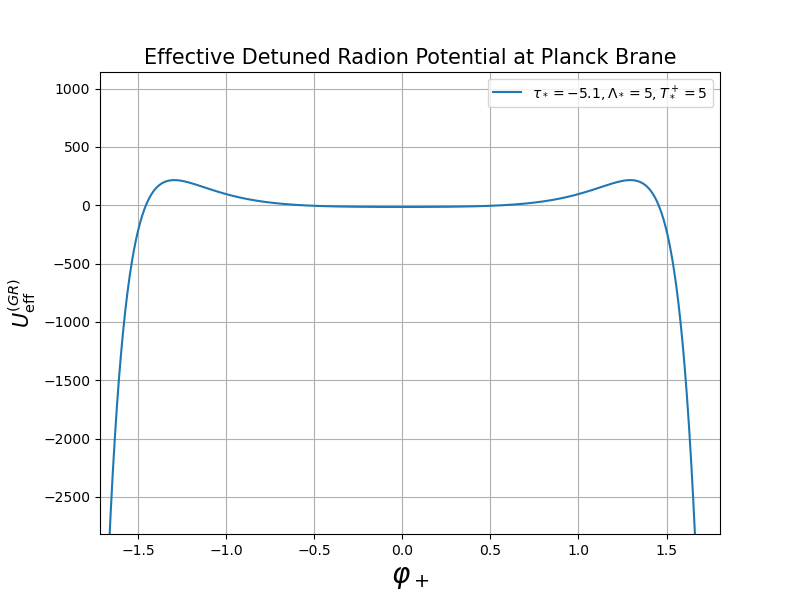}
        \caption{}
    \end{subfigure}
    \hfill
    \begin{subfigure}[b]{0.32\textwidth}
        \centering
        \includegraphics[width=\textwidth]{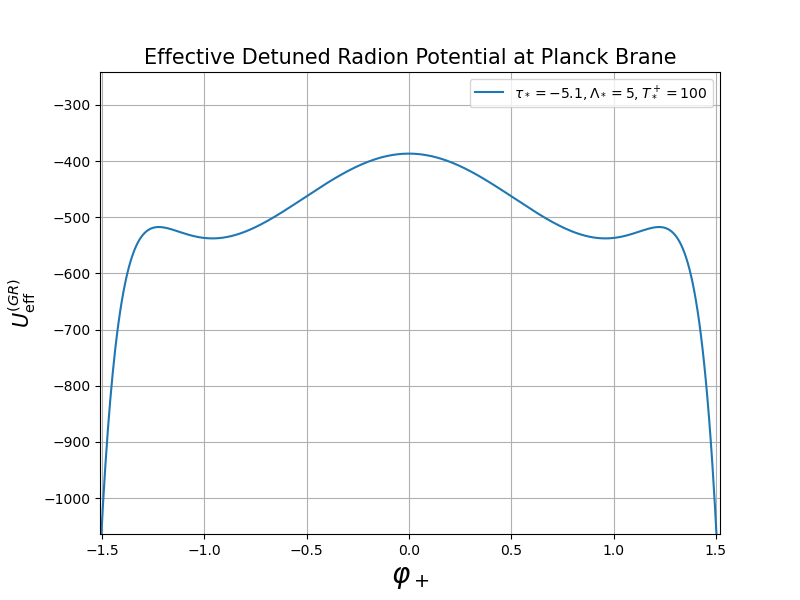}
        \caption{}
    \end{subfigure}
    \hfill
    \begin{subfigure}[b]{0.32\textwidth}
        \centering
        \includegraphics[width=\textwidth]{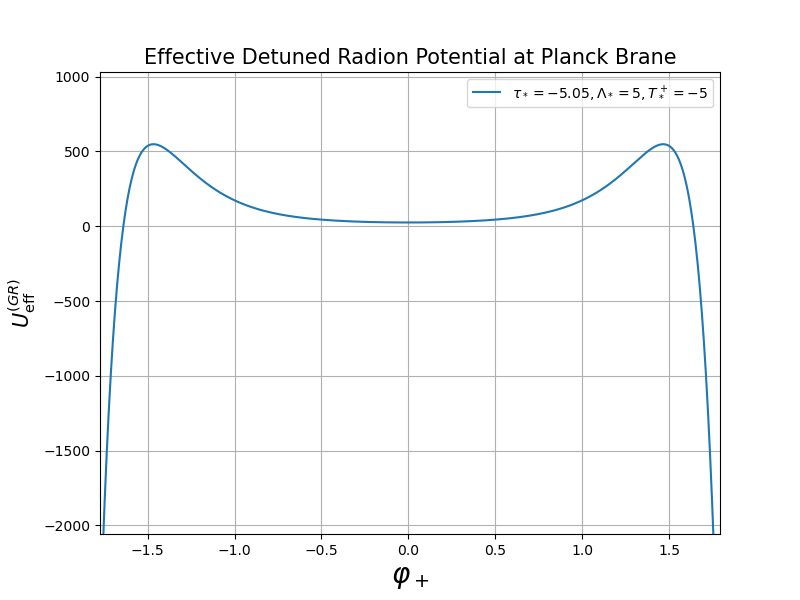}
        \caption{}
    \end{subfigure}
    
    \caption{Plots of the effective detuned radion potential $U_{eff}^{(GR)}$ at Planck brane with respect to the Einstein frame radion field $\varphi_+$ for various parameter choices $\tau_*,\Lambda_*,T_*^+$. We have chosen $\gamma = 16\pi$ and $k = 1$ for obtaining these plots.}
    \label{fig:detuned_planck}
\end{figure}

\noindent And similarly the effective mass squared of the radion field for $s_\pm$ points
\begin{equation}
    \mu_{eff}^2\Bigg|_{\varphi_+(s_\pm)} = \pm \frac{32\pi}{3}s_\pm\sqrt{\tau_*^2 + 2T_*^+(\tau_* + \Lambda_*)}
\end{equation}
Thus if $s_\pm \geq 0$ exists, then $s_+ (s_-)$ gives local minima (maxima) suggesting no tachyonic instability (tachyonic instability) of the radion field at that field configuration. We provide a panel of 6 plots of the effective detuned radion potential $U_{eff}^{(GR)}$ at Planck brane with respect to the Einstein frame radion field $\varphi_+$ for various parameter choices $\tau_*,\Lambda_*,T_*^+$ in Figure (\ref{fig:detuned_planck}).\\

\noindent It is useful to see two limiting cases for the potential given in eq.(\ref{u_p}) in order to explain the plots more extensively. For $\varphi_+\to0$ limit, the potential takes the form $U_{eff}^{(GR)} \sim \text{constant} + \mathcal O(\varphi_+^2)$. Hence the potential is regular at $\varphi_+ = 0$ and moreover $\varphi_+ = 0$ is a stationary point from eq.(\ref{phi_0}). Now for $|\varphi_+| \to \infty$ limit, $U_{eff}^{(GR)}\sim\frac{\Lambda_* + \tau_*}{16}e^{4z|\varphi_+|} +\text{subleading terms}$. That means for $|\varphi_+| \to \infty$, if $\Lambda_* + \tau_* >0$, $U_{eff}^{(GR)}\to\infty$ (bounded) and if $\Lambda_* + \tau_* <0$, $U_{eff}^{(GR)}\to-\infty$ (runaway). And for a specific case $\Lambda_* + \tau_* = 0$ with $|\varphi_+|\to\infty$, we will have $U_{eff}^{(GR)}\sim\frac{\Lambda_*}{2}e^{2z|\varphi_+|}+\text{subleading terms}$. So for $|\varphi_+| \to \infty$, if $\Lambda_*>0$, $U_{eff}^{(GR)}\to\infty$ (bounded) and if $\Lambda_*<0$, $U_{eff}^{(GR)}\to-\infty$ (runaway).\\

\noindent It is indeed true that there are hyperbolic non-linearities present in the potential in eq.(\ref{u_p}). However, this presence of nonlinear hyperbolic terms alone does not guarantee that the instability is quenched, the global behaviour of the potential depends on the parameter choices. In Figure (\ref{fig:detuned_planck}b), (\ref{fig:detuned_planck}d), (\ref{fig:detuned_planck}e), (\ref{fig:detuned_planck}f) are the runaway cases for $|\varphi_+|\to \infty$. And Figure (\ref{fig:detuned_planck}a), (\ref{fig:detuned_planck}c) are the bounded scenario for $|\varphi_+|\to\infty$. This explains the qualitative behaviour observed in Figure (\ref{fig:detuned_planck}). Depending on the parameter choices, the nonlinear hyperbolic terms make either the potential bounded from below or lead to a runaway behaviour at large field values.\\

\noindent The above discussion for Planck brane tells us that for detuned-RS scenario, the tachyonic (in)stability of the radion field at the Planck brane is controlled by the combinations of detuning parameters $\tau_*,\Lambda_*$ with the matter content $T_*^+$.

\subsubsection*{On TeV Brane}

\begin{figure}[h!]
    \centering
    
    \begin{subfigure}[b]{0.32\textwidth}
        \centering
        \includegraphics[width=\textwidth]{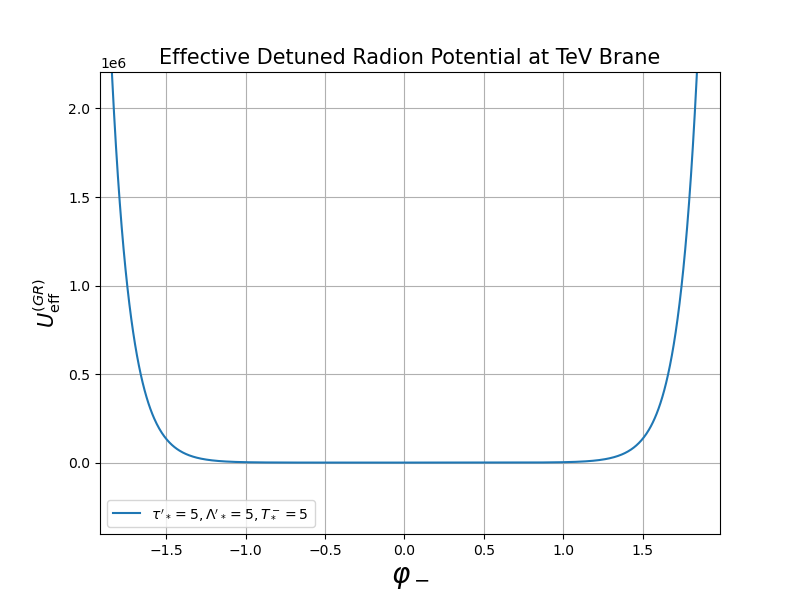}
        \caption{}
    \end{subfigure}
    \hfill
    \begin{subfigure}[b]{0.32\textwidth}
        \centering
        \includegraphics[width=\textwidth]{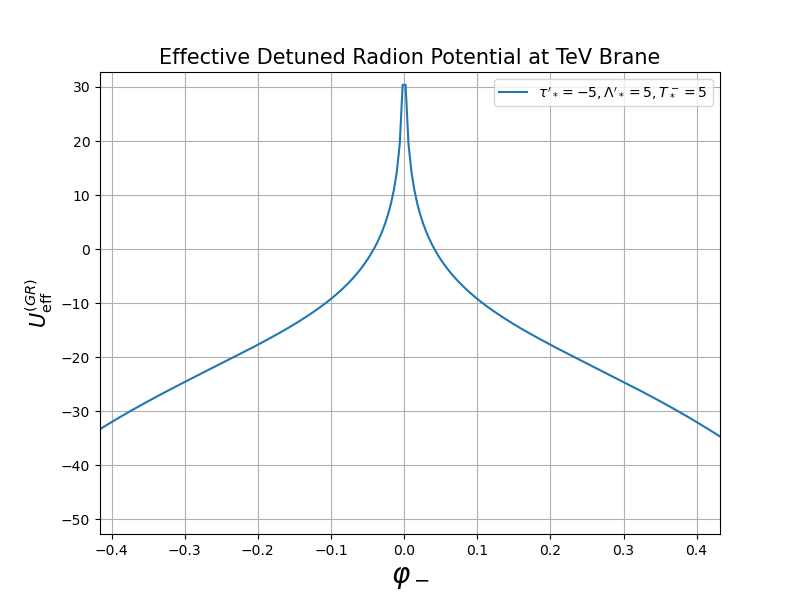}
        \caption{}
    \end{subfigure}
    \hfill
    \begin{subfigure}[b]{0.32\textwidth}
        \centering
        \includegraphics[width=\textwidth]{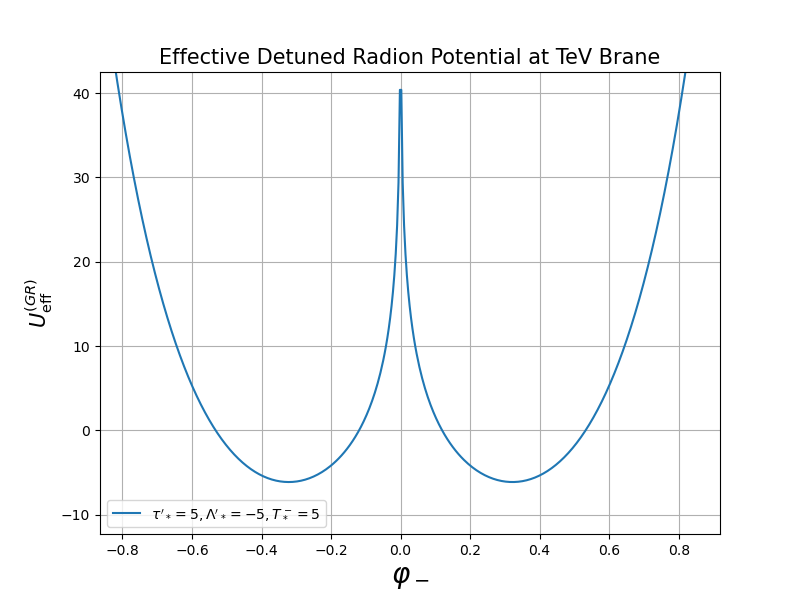}
        \caption{}
    \end{subfigure}
    \\
    \begin{subfigure}[b]{0.32\textwidth}
        \centering
        \includegraphics[width=\textwidth]{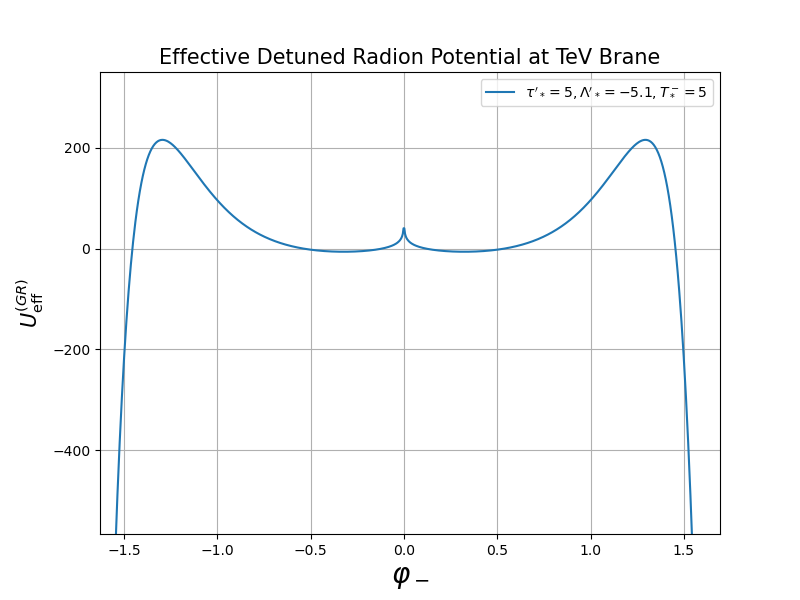}
        \caption{}
    \end{subfigure}
    \hfill
    \begin{subfigure}[b]{0.32\textwidth}
        \centering
        \includegraphics[width=\textwidth]{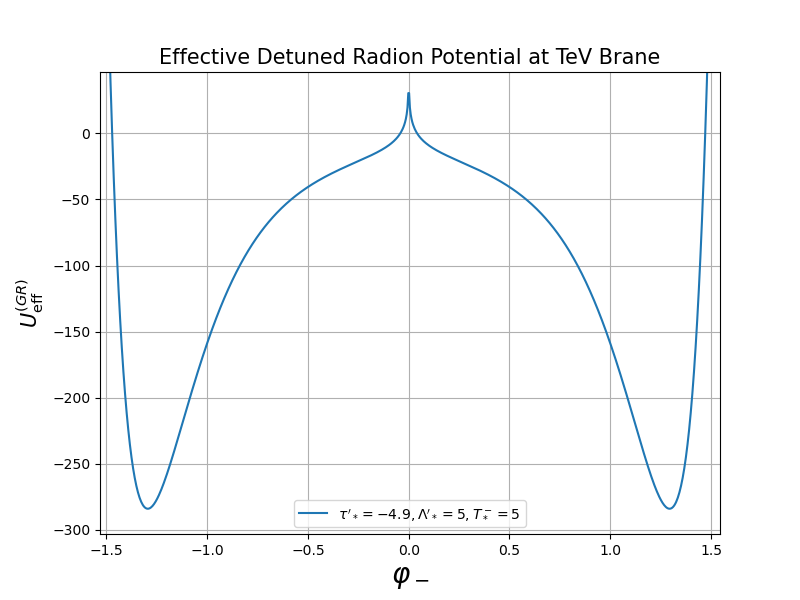}
        \caption{}
    \end{subfigure}
    \hfill
    \begin{subfigure}[b]{0.32\textwidth}
        \centering
        \includegraphics[width=\textwidth]{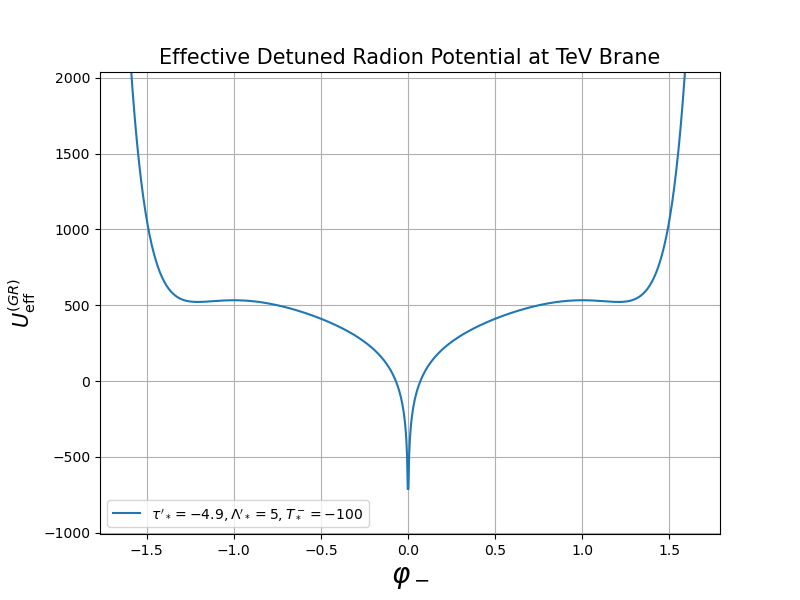}
        \caption{}
    \end{subfigure}
    
    \caption{Plots of the effective detuned radion potential $U_{eff}^{(GR)}$ at TeV brane with respect to the Einstein frame radion field $\varphi_-$ for various parameter choices $\tau'_*,\Lambda'_*,T_*^-$. We have chosen $\gamma = 16\pi$ and $k = 1$ for obtaining these plots.}
    \label{fig:detuned_tev}
\end{figure}

Similar to the Planck brane analysis, again using eq.(\ref{22}) we can write the radion potential in term of the BD scalar $\Phi_-$ on the TeV brane as
\begin{equation}\label{47}
    V_{GR}(\Phi_-) = \frac{k^4}{A^4}\left[\Lambda_{4D}' + \tau'\left(1+ \frac{\Phi_-}{\gamma}\right)^2\right]
\end{equation}
The corresponding Einstein frame potential is given by
\begin{equation}\label{48}
    U_{GR}(\varphi_-) \equiv \frac{16\pi}{\Phi_-^2}V_{GR}(\Phi_-)
\end{equation}
We can express the BD scalar $\Phi_-$ in terms of the Einstein frame radion field $\varphi_-$ as~\footnote{see Appendix \ref{A} for the derivation}
\begin{equation}\label{a61}
    \Phi_- = \gamma\;\text{csch}^2\left(\sqrt{\frac{4\pi}{3}}\varphi_-\right)
\end{equation}
Then the complete effective potential of radion field at the TeV brane in terms of Einstein frame radion field $\varphi_-$ is given by
\begin{equation}\label{u_m}
    U_{eff}^{(GR)}(\varphi_-) = \Lambda'_*\sinh^4\left(\sqrt{\frac{4\pi}{3}}\varphi_-\right) + \tau'_*\cosh^4\left(\sqrt{\frac{4\pi}{3}}\varphi_-\right) + T^-_*\ln\text{csch}^2\left(\sqrt{\frac{4\pi}{3}}\varphi_-\right) - T_*^-\ln\gamma
\end{equation}
where, $\tau'_* \equiv 16\pi k^4\tau'/A^4\gamma^2,\;\; \Lambda'_* \equiv 16\pi k^4\Lambda'_{4D}/A^4\gamma^2,\;\;T_*^- \equiv 8\pi T_{(-)}$.
Again by doing the similar analysis as in section \ref{ST}, that is, linearize the Einstein frame radion field at first order around some $\varphi_-$, we get the effective mass squared for the first order scalar perturbation $\mu_{eff}^2 = U^{(GR)''}_{eff}$. Assuming $z = \sqrt{\frac{4\pi}{3}}\varphi_-$ and $s \equiv \sinh^2z \geq 0$, we write the effective potential in terms of $s$
\begin{equation}
    U_{eff}^{(GR)} = (\Lambda'_* + \tau'_*)s^2 + 2\tau'_*s + \tau'_* - T_*^-\ln s - T_*^-\ln\gamma
\end{equation}
Here, $U_{eff}^{(GR)'} = 0$ does have solutions. The $U_{eff}^{(GR)'}$ is given as
\begin{align}
U_{eff}^{(GR)'}
&= \sqrt{\frac{4\pi}{3}}\frac{1}{s}2\sinh z\cosh z\left[2(\Lambda'_*+\tau'_*)s^2 + 2\tau'_*s -T_*^-\right]
\end{align}
Clearly, $U_{eff}^{(GR)'} = 0$ solutions are
\begin{equation}
    s_\pm = \frac{-\tau'_* \pm \sqrt{\tau'^2_* + 2T_*^-(\Lambda_*' + \tau_*')}}{2(\Lambda'_* + \tau_*')} \implies \varphi_- = \pm \sqrt{\frac{3}{4\pi}}\sinh^{-1}\sqrt{s_\pm}
\end{equation}
The last solution is valid only when $s_\pm \geq 0$, $\Lambda'_*+\tau'_* \neq 0$ and $\tau_*^{'2}+2T_*^-(\Lambda'_*+\tau'_*) \geq 0$. Note that there is no stationary point at $\varphi_- = 0$ unlike Planck brane. The effective mass squared of the Einstein frame radion field at TeV brane at the point $s_\pm$ is given by
\begin{equation}
    \mu_{eff}^2\Bigg|_{\varphi_-(s_\pm)} = \frac{16\pi}{3}s_\pm(1+s_\pm)\left[2(\Lambda'_* + \tau'_*) + \frac{T_*^-}{s_\pm^2}\right]
\end{equation}
The sign of $\mu_{eff}^2$ depends on the sign of the quantity in the square brackets. We provide a panel of 6 plots of the effective detuned radion potential $U_{eff}^{(GR)}$ at TeV brane with respect to the Einstein frame radion field $\varphi_-$ for various parameter choices $\tau'_*,\Lambda'_*,T_*^-$ in Figure (\ref{fig:detuned_tev}). \\

\noindent Similar to Planck brane scenario, it is useful to see two limiting cases for the potential given in eq.(\ref{u_m}) in order to explain the plots more extensively. For $\varphi_-\to0$ limit, the potential takes the form $U_{eff}^{(GR)} \sim \text{constant} - 2T_*^-\ln|\varphi_-| + \mathcal O(\varphi_-^2)$. Hence there is a logarithmic singularity at $\varphi_- = 0$, which is basically arising from the $\ln\text{csch}^2$ term in eq.(\ref{u_m}). Then for $\varphi_-\to0$, $U_{eff}^{(GR)}\to\infty$ for $T_*^->0$ and $U_{eff}^{(GR)}\to-\infty$ for $T_*^-<0$. Now for $|\varphi_-| \to \infty$ limit, $U_{eff}^{(GR)}\sim\frac{\Lambda_*' + \tau_*'}{16}e^{4z|\varphi_-|} +\text{subleading terms}$. That means for $|\varphi_-| \to \infty$, if $\Lambda_*' + \tau_*' >0$, $U_{eff}^{(GR)}\to\infty$ (bounded) and if $\Lambda_*' + \tau_*' <0$, $U_{eff}^{(GR)}\to-\infty$ (runaway). And for a specific case $\Lambda_*' + \tau_*' = 0$ with $|\varphi_-|\to\infty$, we will have $U_{eff}^{(GR)}\sim-\frac{\Lambda_*'}{2}e^{2z|\varphi_-|}+\text{subleading terms}$. So for $|\varphi_-| \to \infty$ if $\Lambda_*'<0$, $U_{eff}^{(GR)}\to\infty$ (bounded) and if $\Lambda_*'>0$, $U_{eff}^{(GR)}\to-\infty$ (runaway).\\

\noindent It is indeed true that there are hyperbolic non-linearities present in the potential in eq.(\ref{u_m}). However, this presence of nonlinear hyperbolic terms alone does not guarantee that the instability is quenched, the global behaviour of the potential depends on the parameter choices. All the figures in Figure (\ref{fig:detuned_tev}) have a logarithmic singularity at $\varphi_-=0$ (the logarithmic singularity is not clearly visible in Figure (\ref{fig:detuned_tev}a) because of the chosen vertical-axis range). In Figure (\ref{fig:detuned_tev}b), (\ref{fig:detuned_tev}d) are the runaway cases for $|\varphi_-|\to \infty$. And Figure (\ref{fig:detuned_tev}a), (\ref{fig:detuned_tev}c), (\ref{fig:detuned_tev}e), (\ref{fig:detuned_tev}f) are the bounded scenario for $|\varphi_-|\to\infty$. This explains the qualitative behaviour observed in Figure (\ref{fig:detuned_tev}). Depending on the parameter choices, the nonlinear hyperbolic terms make either the potential bounded from below or lead to a runaway behaviour at large field values.\\

\noindent We can see that the Einstein frame radion field $\varphi_- \to 0$ corresponds to the Jordan frame radion field $\Phi_- \to \infty$. Now recall the conformal transformation in eq.(\ref{28}). So exactly at $\varphi_- = 0$ the conformal transformation itself is no longer defined. And in that case, using eq.(\ref{22}) and (\ref{a61}), one can see $T(x) \to \infty$. Physically, the separation between the Planck and TeV branes becomes infinite, pushing the TeV brane to the AdS horizon and it entirely decouples from the 4D effective theory. \\

\noindent The above discussion for TeV brane tells us that for detuned-RS scenario, the tachyonic (in)stability of the radion field at the TeV brane is controlled by the combinations of detuning parameters $\tau'_*,\Lambda'_*$ with the matter content $T_*^-$.\\

\noindent In detuned-RS scenario, there is a possibility of spontaneous scalarization of compact objects ($T<0$) in Planck brane as well, unlike the tuned-RS case, for some combination of the brane detuning parameters. Therefore in detuned-RS, the spontaneous scalarization process may be feasible in both the Planck and TeV brane if some of the above discussed criteria are fulfilled. In principle, one can play with the detuning parameters and generate different types of effective potentials. For some of the potentials, there is a catastrophic tachyonic instability of the detuned radion field and for some of the potentials, there are unstable maxima along with some stable minima.\\

\noindent In the following subsection, we will discuss the radion-stabilized RS scenario and deduce some important conclusions for tachyonic instability of the \textit{stabilized} radion field on either brane.

\subsection{Goldberger-Wise Stabilized RS}\label{GW}

We again set the fine-tuning conditions eq.(\ref{19}), which make the gravitational potential term vanish. However, then the radion field could take up any possible value. To fix this, one needs a stabilization mechanism that fixes the interbrane separation and makes the radion acquire a mass. Such mechanism was proposed by Goldberger and Wise~\cite{23}. In this construction, a massive 5D field $\Phi$ is sourced at the boundaries and acquires a VEV whose value depends on the location in the extra dimension. After integrating over the extra dimension, this generates a radion potential in the low energy effective theory. In the original GW construction, the quartic potential for the radion from the gravity sector was still kept tuned to zero and only the dynamics of the scalar field $\Phi$ contributed to the radion potential.\\

\noindent We add to the original RS action a scalar field $\Phi$ with the following bulk action
\begin{equation}\label{52}
    S_{bulk} = \int d^4x\int_0^{r_c}dz\sqrt{G}\;\left(G^{AB}\partial_A\Phi\partial_B\Phi - m^2\Phi^2\right)
\end{equation}
where $G_{AB}$ with $A,B = \mu,z$ is given by eq.(\ref{18}) and $m$ is the mass of the bulk scalar field $\Phi$. We also include interaction terms on the hidden and visible branes (at $z=0$ and $z=r_c$ respectively) given by
\begin{align}\label{53}
S_{hid} &= -\int d^4x \sqrt{-g_h}\;\lambda_h(\Phi^2-v_h^2)^2\nonumber \\
S_{vis} &= -\int d^4x \sqrt{-g_v}\;\lambda_v(\Phi^2-v_v^2)^2
\end{align}
The terms on the branes cause $\Phi$ to develop a $z$-dependent vacuum expectation value of $\Phi(z)$ which is determined classically by solving the differential equation
\begin{equation}
    0 = -\partial_z\left(e^{-4k|z|}\partial_z\Phi\right) + m^2e^{-4k|z|}\Phi + 4e^{-4k|z|}\lambda_h\Phi(\Phi^2 - v_h^2)\delta(z) + 4e^{-4k|z|}\lambda_v\Phi(\Phi^2 - v_v^2)\delta(z-r_c) 
\end{equation}
One can solve this differential equation with proper boundary conditions and plugging this solution back into the bulk scalar field action in eq.(\ref{52}) and integrating over $z$ yields an effective 4D potential for $T(x)$. Now suppose that $m/k \ll 1$ and so $m^2/4k^2$ is a small quantity. In the large $kr_c$ limit on the Planck brane, the effective 4D radion potential is given by
\begin{equation}\label{54}
    V_{GW}(\phi) = \frac{k}{A^4}\phi^4\left[v_h - v_v \left(\frac{\phi}{A}\right)^\epsilon\right]^2
\end{equation}
where $\epsilon = m^2/4k^2$ and $\phi = Ae^{-kr_cT}$ with $A^2 = 12M_*^3/k$. To obtain the TeV brane potential, we need to multiply the Planck brane potential by $e^{4kr_cT}$. Ignoring the terms proportional to $\epsilon$, this potential has a minimum at
\begin{equation}\label{55}
    kr_cT = \frac{4k^2}{m^2}\ln\left(\frac{v_h}{v_v}\right)
\end{equation}
Using $v_h/v_v = 1.5$ and $m/k = 0.2$ in eq.(\ref{55}) yields $kr_cT \simeq 36$ and as one can see, no unnatural fine-tuning of parameters is required to solve the hierarchy problem. Since the modulus gets stabilized to a fixed value, we can clearly see that the Brans-Dicke fields on both the branes which depend on the modulus field become constants. As a result, the theory on either brane eventually turns out to be GR.\\

\noindent Similar to the previous sections, we do the first order scalar perturbation around some $\varphi_\pm$ and the effective mass squared of the Einstein frame radion field is
\begin{equation}\label{56}
    \mu_{eff}^2 = U''_{GW}(\varphi_{\pm}) - 16\pi \alpha'(\varphi_{\pm})T_{(\pm)} \equiv U_{eff}^{(GW)''}(\varphi_\pm)
\end{equation}

\noindent Below we present the evaluation of effective mass squared of radion field about the minimum of the GW radion potential on the either branes separately.

\subsubsection*{On Planck Brane}

On the Planck brane, the GW radion potential can be written in terms of BD scalar $\Phi_+$ using the eq.(\ref{26}) as
\begin{equation}\label{57}
    V_{GW}(\Phi_+) = k\left(1 - \frac{\Phi_+}{\gamma}\right)^2\left[v_h - v_v\left(1 - \frac{\Phi_+}{\gamma}\right)^{\epsilon/2}\right]^2
\end{equation}
The corresponding Einstein frame potential is given by
\begin{equation}\label{43}
    U_{GW}(\varphi_+) \equiv \frac{16\pi}{\Phi_+^2}V_{GW}(\Phi_+)
\end{equation}
The complete GW effective potential of the radion field at the Planck brane is given by (using eq.(\ref{a51}))
\begin{equation}
    U_{eff}^{(GW)}(\varphi_+) = \frac{16\pi k}{\gamma^2}\sinh^4\left(\sqrt{\frac{4\pi}{3}}\varphi_+\right)\left[v_h - v_v\Big\{\tanh\left(\sqrt{\frac{4\pi}{3}}\varphi_+\right)\Big\}^\epsilon\right]^2 + T^+_*\ln\text{sech}^2\left(\sqrt{\frac{4\pi}{3}}\varphi_+\right) - T_*^+\ln\gamma
\end{equation}
The standard GW minimum is obtained by setting $T_*^+ = 0$ and the non-trivial minimum lies at $\varphi_+ \sim 10^{-16} \ll1$, which is almost numerically indistinguishable from the origin $\varphi_+ = 0$.\\

\begin{figure}[h!]
    \centering
    \includegraphics[width=0.7\linewidth]{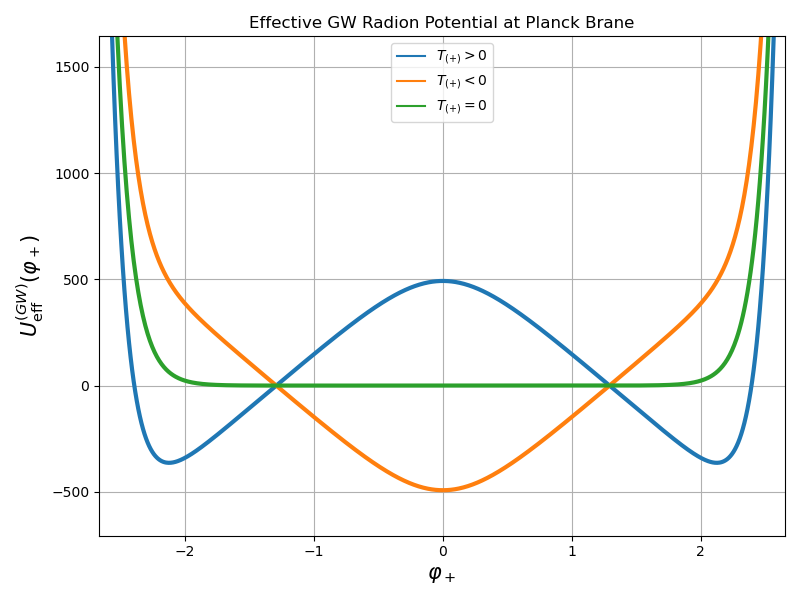}
    \caption{Plot between the effective GW radion potential at the Planck brane and the Einstein frame radion field $\varphi_+$ for three scenarios - $T_{(+)} = 0$, $T_{(+)} > 0$ and $T_{(+)} < 0$. We have chosen $v_h/v_v = 1.5$, $m/k = 0.2$, $\gamma = 16\pi$ and $k = 1$ for obtaining these plots.}
    \label{fig:gw_planck}
\end{figure}

\noindent Assuming $a = \sqrt{4\pi/3}$, $z = a\varphi_+$, $B(z) = v_h - v_v\tanh^\epsilon z$ and $C = 16\pi k/\gamma^2$, we write the effective GW potential on the Planck brane as follows
\begin{equation}
    U_{eff}^{(GW)}(\varphi_+) = C\sinh^4zB^2(z) + T_*^+\ln\text{sech}^2z - T_*^+\ln\gamma
\end{equation}
The first derivative of the effective potential with respect to $\varphi_+$
\begin{equation}
    U_{eff}^{(GW)'} = a\left[C\left(4\sinh^3z\cosh z B^2(z) - 2\epsilon v_v\sinh^4zB(z)\tanh^{\epsilon - 1}z\;\text{sech}^2z\right) - 2T_*^+\tanh z\right]
\end{equation}
Clearly, $U_{eff}^{(GW)'} = 0$ does have the following solutions
\begin{equation}
    z = 0 \implies \varphi_+ = 0
\end{equation}
and
\begin{equation}
    2T_*^+ = C\cosh z\left(4\sinh^2z\cosh z B^2(z) - 2\epsilon v_v\sinh^3zB(z)\tanh^{\epsilon - 1}z\;\text{sech}^2z\right)
\end{equation}
The last solution is a transcendental equation which can only be solved numerically. But we will not do that. We will visualize these stationary points from the plots. Now we will see the effective mass squared of the radion field at $\varphi_+ = 0$, considering only up to quadratic order field terms
\begin{equation}
    \mu^2_{eff}\Bigg|_{\varphi_+ = 0} \simeq -\frac{8\pi}{3}T_*^+
\end{equation}
Thus, for $T_*^+ < 0$, $\mu_{eff}^2 > 0$, means no tachyonic instability of the radion field (a local minimum) and for $T_*^+ > 0$, $\mu_{eff}^2 < 0$, means radion field is prone to tachyonic instability (a local maximum) at $\varphi_+ = 0$. While for $T_*^+ = 0$, $\mu_{eff}^2 \simeq 0$, means the effective potential is flat (up to quadratic approximation) at $\varphi_+ = 0$ (no tachyonic instability). We present the plot between the effective GW radion potential at the Planck brane and the Einstein frame radion field $\varphi_+$ in Figure (\ref{fig:gw_planck}).

\subsubsection*{On TeV Brane}

\begin{figure}[h!]
    \centering
    \includegraphics[width=0.7\linewidth]{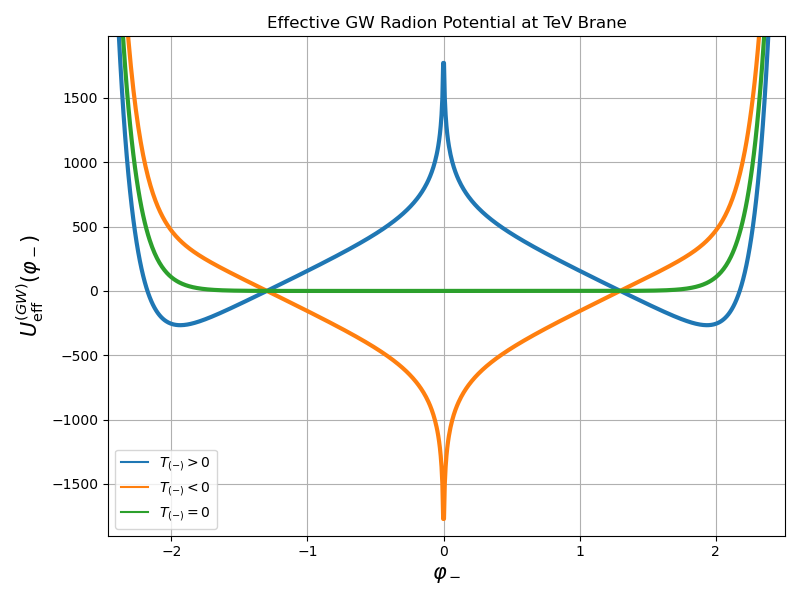}
    \caption{Plot between the effective GW radion potential at the TeV brane and the Einstein frame radion field $\varphi_-$ for three scenarios - $T_{(-)} = 0$, $T_{(-)} > 0$ and $T_{(-)} < 0$. We have chosen $v_h/v_v = 1.5$, $m/k = 0.2$, $\gamma = 16\pi$ and $k = 1$ for obtaining these plots.}
    \label{fig:gw_tev}
\end{figure}

On the TeV brane, the GW radion potential can be written in terms of BD scalar $\Phi_-$ using the eq.(\ref{26}) as
\begin{equation}\label{61}
    V_{GW}(\Phi_-) = k\left[v_h - v_v\left(1 + \frac{\Phi_-}{\gamma}\right)^{-\epsilon/2}\right]^2
\end{equation}
The corresponding Einstein frame potential is given by
\begin{equation}\label{43}
    U_{GW}(\varphi_-) \equiv \frac{16\pi}{\Phi_-^2}V_{GW}(\Phi_-)
\end{equation}
The complete GW effective potential of the radion field at the TeV brane is given by (using eq.(\ref{a51}))
\begin{equation}
    U_{eff}^{(GW)}(\varphi_-) = \frac{16\pi k}{\gamma^2}\sinh^4\left(\sqrt{\frac{4\pi}{3}}\varphi_-\right)\left[v_h - v_v\Big\{\tanh\left(\sqrt{\frac{4\pi}{3}}\varphi_-\right)\Big\}^\epsilon\right]^2 + T^-_*\ln\text{csch}^2\left(\sqrt{\frac{4\pi}{3}}\varphi_-\right) - T_*^-\ln\gamma
\end{equation}
Again, assuming $a = \sqrt{4\pi/3}$, $z = a\varphi_-$, $B(z) = v_h - v_v\tanh^\epsilon z$ and $C = 16\pi k/\gamma^2$, we write the effective GW potential on the Planck brane as follows
\begin{equation}
    U_{eff}^{(GW)}(\varphi_-) = C\sinh^4zB^2(z) + T_*^-\ln\text{csch}^2z - T_*^-\ln\gamma
\end{equation}
The first derivative of the effective potential with respect to $\varphi_-$
\begin{equation}
    U_{eff}^{(GW)'} = a\left[C\left(4\sinh^3z\cosh z B^2(z) - 2\epsilon v_v\sinh^4zB(z)\tanh^{\epsilon - 1}z\;\text{sech}^2z\right) - 2T_*^-\coth z\right]
\end{equation}
Clearly, $U_{eff}^{(GW)'} = 0$ does have the following solutions
\begin{equation}
    2T_*^-\coth z = C\left(4\sinh^3z\cosh z B^2(z) - 2\epsilon v_v\sinh^4zB(z)\tanh^{\epsilon - 1}z\;\text{sech}^2z\right)
\end{equation}
The above solution is a transcendental equation which can only be solved numerically. But we will not do that. We will visualize these stationary points from the plot. Note that $\varphi_- = 0$ is no longer an extremum point for $T_-^* \neq 0$, rather the effective potential becomes singular at this point ($\varphi_- = 0$ is the minima for only $T_-^* = 0$, the standard GW scenario), similar logarithmic singularity we have seen in section \ref{detuned-RS}. We present the plot between the effective GW radion potential at the TeV brane and the Einstein frame radion field $\varphi_-$ in Figure (\ref{fig:gw_tev}).
\\

\noindent In radion-stabilized scenario, there is no favor to occur scalarization of compact objects ($T < 0$) in both TeV brane and Planck brane. For both cases, matter with $T<0$, increases the stability of the GW minima. Although matter with positive trace of energy-momentum tensor ($T>0$) can trigger the tachyonic instability of radion field on the either branes which is sitting at the GW minimum. We know the bulk theory is of pure Einstein theory, but the effective on-brane theory is of scalar-tensor nature (see from eq.(\ref{21})). In the absence of any on-brane matter fields, the radion field is stabilized via GW mechanism and sitting at the GW minimum, a constant scalar field configuration, which hints at the pure Einstein gravity (GR) on the branes. But in the presence of matter ($T>0$), the radion field on the either brane gains tachyonic instability and due to the non-linear terms in the effective potential, it rolls down to one of the minima. Thus the on-brane gravity drives away from GR. \\

\noindent Basically, introducing $T>0$ matter on both Planck and TeV brane effectively changes the VEV of the radion field. Usually, in the absence of matter, we fine tune the ratio $v_h/v_v = 1.5$ and the mass of the stabilizing field $m/k = 0.2$ to obtain $kr_cT \simeq 36$, which can solve the gauge hierarchy problem. But introducing $T>0$ matter on the branes changes this VEV and with the same parameter choices, $kr_cT$ value is very much off from $36$. Hence, in the presence of matter ($T>0$) radion field can be stabilized, but it does not solve the gauge hierarchy problem. For example : in Planck brane with $T_{(+)} = 5 > 0$ and $v_h/v_v = 1.5$, $m/k = 0.2$ and $\gamma = 16\pi$, one of the minima occurs at $\varphi_+ \simeq 2.27994$, which in turn gives $kr_cT \simeq 17 \times 10^{-5}$, which is off by 5 orders of magnitude! Thus the resolution of the gauge hierarchy problem is completely destroyed. \\

\noindent In the next subsection, we present the effective mass squared of the radion field in two specific on-brane geometry (dS$_4$ and AdS$_4$), as an example of the fact that we need to specify the background geometry in order to obtain exact tachyonic instability condition of the radion field.

\subsection{dS$_4$ and AdS$_4$ Geometry on RS Branes}\label{flrw}

Till now we have not considered any background geometry on the RS branes. And hence the tachyonic (in)stability criteria stated in the previous subsections (\ref{Tuned-RS}, \ref{detuned-RS} and \ref{GW}) of the radion field was basically a \textit{weaker condition} (also discussed in section \ref{curved}) of tachyonic instability. It is very much instructive to consider the maximally symmetric spacetime solutions: de-Sitter (dS) spacetime and Anti de-Sitter (AdS) spacetime on the RS branes. The motivation for considering the dS$_4$ brane is the inflationary or dark-energy dominated epoch of cosmology, where the Hubble parameter $H_{dS}$ is a constant. And the motivation behind the AdS$_4$ brane is due to the string theory connection of the braneworld scenarios (braneworld theories are the low-energy effective theories coming from fundamental string theories which live in AdS spacetimes; for example see Ref.~\cite{KS,KS2}). \\

\subsubsection{dS$_4$ Brane}

\noindent For constructing a dS$_4$ brane, it is sufficient to consider spatially homogeneous, isotropic and time-dependent Friedmann-Lemaitre-Robertson-Walker (FLRW) geometry on the RS branes. FLRW geometry on the RS branes enables us to study the cosmological evolution of the brane matter, radiation or vacuum energy. In such time-dependent scenario, the radion field also evolve in cosmic time and influence the cosmic expansion rate. It gives various opportunity to study its behavior and stabilization in different epochs of cosmological evolution. Hence we consider the spatially flat FLRW geometry on the either branes
\begin{equation}\label{64}
    ds^2 = e^{-2kT(x)|z|}ds^2_{brane} + T^2(x)dz^2
\end{equation}
with
\begin{equation}\label{brane}
    ds^2_{brane} = -dt^2 + a^2(t)\delta_{ij}dx^idx^j
\end{equation}
where $a(t)$ is the scale factor. For simplicity, we only present the analysis on the Planck brane. The TeV brane analysis will be similar as the TeV brane is conformally related to Planck brane. On Planck brane, the Einstein frame metric ($\tilde g$) is related to the Jordan frame metric ($g$) as $\tilde g_{\mu\nu(+)} = \Phi_+g_{\mu\nu}$ and $\sqrt{-\tilde g} = \Phi_+^2a^3(t)$. If we consider the radion field to be spatially homogeneous (zero-momentum mode $k=0$), then its equation of motion in FLRW background can be expressed as
\begin{equation}\label{65}
    -\frac{1}{\Phi_+}\left[\ddot\varphi_+ + \left(3H + \frac{\dot\Phi_+}{\Phi_+}\right)\dot\varphi_+\right] = U_i'(\varphi_+) - 16\pi\alpha(\varphi_+)T_{(+)}
\end{equation}
where $i$ corresponds to three cases - Tuned-RS ($U_i \equiv 0$), Detuned-RS ($U_i \equiv U_{GR}$), GW Stabilized RS ($U_i \equiv U_{GW}$). Here, $H \equiv \dot a/a$ is the Hubble parameter on the Planck brane. Again we will do the first order scalar perturbation around some $\varphi_+ = \varphi_{0+}$ with $\alpha(\varphi_{0+}) = U'_i(\varphi_{0+}) = 0$ in order to obtain the effective mass squared of the radion field on the Planck brane with FLRW background geometry. The first order perturbed equation is
\begin{equation}\label{66}
    \delta\ddot\varphi_+ + 3H\delta\dot\varphi_+ + \mu_{eff}^2\delta\varphi_+ = 0
\end{equation}
with the effective mass squared of the radion field expressed as
\begin{equation}\label{67}
    \mu_{eff}^2 = \Phi_{0+}\left[U''_i(\varphi_{0+}) - 16\pi\alpha'(\varphi_{0+})T_{(+)}\right]
\end{equation}
Assume the change of variable $\delta\varphi_+ = a^{-3/2}u$ in order to write the eq.(\ref{66}) in more familiar form : $\ddot u + \Omega^2(t)u = 0$ where
\begin{equation}\label{68}
    \Omega^2(t) = \mu_{eff}^2 - \frac{3}{2}\dot H - \frac{9}{4}H^2
\end{equation}
If $\Omega^2<0$, then $u \sim e^{\Omega t}$ and the perturbation grows exponentially, which hints at the tachyonic instability of the radion field on the Planck brane with FLRW geometry on the brane. Hence the \textit{exact} condition for tachyonic instability of the radion field on the Planck brane is given by
\begin{equation}\label{69}
    \mu_{eff}^2 < \frac{3}{2}\dot H + \frac{9}{4}H^2
\end{equation}
For de-Sitter (dS$_4$) Planck brane scenario, the scale factor $a(t) = e^{H_{dS}t}$ with the Hubble parameter $H_{dS}$ being constant and $\dot H_{dS} = 0$. Hence the effective mass squared of radion field is given by
\begin{equation}\label{dS_eq}
    \mu_{eff}^2 < \frac{9}{4}H_{dS}^2\;\;\;\;:\;\;\text{for de-Sitter Brane}
\end{equation}
for tachyonic instability to occur. This beautifully explains the fact that the exact condition for tachyonic instability of radion field on the brane (here, Planck brane) is geometry dependent with the Hubble friction playing an important role.

\subsubsection{AdS$_4$ Brane}

To describe Anti de-Sitter brane, the spatially flat metric in eq.(\ref{brane}) is not sufficient. One needs an open hyperbolic geometry in order to describe AdS spacetimes. But we can get the similar equation for AdS$_4$ brane as of eq.(\ref{dS_eq}) from an easy way. The on-brane Ricci scalar for the metric in eq.(\ref{brane}) is given as
\begin{equation}
    R_{brane} = 6(\dot H + 2H^2)
\end{equation}
Then for dS$_4$ brane, $\dot H_{dS} = 0$ and we get $R_{brane} = 12H^2_{dS}$. From eq.(\ref{dS_eq}), we can write a covariant tachyonic instability condition in terms of $R_{brane}$ as
\begin{equation}\label{cov}
    \mu^2_{eff} < \frac{3}{16}R_{brane}
\end{equation}
AdS$_4$ brane is a 4D spacetime with a constant negative curvature, that is
\begin{equation}
    R_{brane} = -\frac{12}{L_{AdS}^2}
\end{equation}
where, $L_{AdS}$ is called the AdS curvature radius. Therefore from the covariant relation in eq.(\ref{cov}), we get the tachyonic instability condition for AdS brane
\begin{equation}
    \mu_{eff}^2 < - \frac{9}{4L_{AdS}^2}\;\;\;\;:\;\;\text{for Anti de-Sitter Brane}
\end{equation}
This is the well-known \textit{Breitenlohner-Freedman} bound~\cite{BF1,BF2} for AdS spacetimes, demonstrating that the radion field enjoys an enhanced stability region in negatively curved spacetimes where it can possess a negative mass squared without triggering tachyonic instability. The AdS$_4$ branes are also known as the \textit{Karch-Randall} (KR) braneworlds~\cite{KR1,KR2} which lives in warped AdS$_5$ bulk. There is a recent growing interest in KR braneworlds, which is driven by their applications to localized massive gravity, AdS/CFT correspondence~\cite{Geng1} and black-hole information paradox~\cite{Geng2}.

\section{Conclusions and Outlooks}\label{conclusion}

\subsection{Conclusions}

We have reviewed the tachyonic instability of scalar field in both flat and curved spacetime scenarios. Then we have derived the modified field equations of a scalar-tensor theory in Einstein frame. We found the effective mass squared of the scalar degree of freedom of scalar-tensor theories and $f(R)$ theories using first order scalar perturbations. We have provided a compact way to write the effective on-brane 4D action in both Jordan and Einstein frame. With the same techniques developed for scalar-tensor theories, we discussed the effective potential in the presence of matter and the corresponding effective mass squared of the radion field under tuned-RS, detuned-RS and GW-stabilized RS scenarios. Then we dicussed different possibilities of tachyonic instability and spontaneous scalarization in three different scenarios (tuned-RS, detuned-RS and GW-stabilized RS) on the Planck and TeV branes separately. We have found that the spontaneous scalarization of compact objects will be more feasible in TeV brane rather than Planck brane with the fine-tunings conditions. For detuned-RS scenarios, we have found that there is a possibility of spontaneous scalarization of compact objects in both Planck brane as well as in TeV brane for some specific combination of the brane detuning parameters and matter trace. The original Goldberger-Wise mechanism was formulated without any matter content on the either RS branes. But if we introduce on-brane matter fields, the stabilization of the radion field may get destabilized. We have found that (see section \ref{GW}) for matter fields with the trace of energy momentum tensor $T>0$ will destabilize the radion field at the GW minimum and drives it towards a stable minima (with the help of non-linearities present in the theory), changing the VEV of radion. This change in VEV completely destroys the resolution of the gauge hierarchy problem. We have also found that matter with $T<0$ does not alter the stabilization and the VEV of the radion field at the GW minimum, indicating no favor towards the spontaneous scalarization of compact objects ($T<0$) on the branes. Finally, we talked about dS and AdS branes and we derived the exact conditions for tachyonic instability of the radion field for both dS and AdS branes.

\subsection{Future Outlooks}

\noindent We are going to discuss a possible future direction aligned with this work. There is an interesting dual connection with the 4D effective theory on the RS branes with the 4D on-brane $f(R)$ gravity theory at a specific epoch of cosmology - that is, inflation. The 4D effective scalar degree of freedom coming from a higher dimensional theory, that is, the radion field can have a action-level dual description as a $f(R)$ scalar degree of freedom in inflationary cosmology with the radion playing the role of slow roll inflaton field on the brane. Recall the action in eq.(\ref{21}), the 4D effective on-brane action in Jordan frame with the RS fine-tuning conditions which does not contain any radion potential. But for detuned-RS and GW stabilized RS, we do get a corresponding radion potential. Considering such a radion potential we rewrite the on-brane RS action as
\begin{equation}\label{80}
    S = \int d^4x \sqrt{-g_{(\pm)}}\frac{1}{16\pi} \left(\Phi_{\pm}R - \frac{\omega_\pm(\Phi_\pm)}{\Phi_\pm}\nabla_\mu\Phi_\pm\nabla^\mu\Phi_\pm - V_i(\Phi_\pm)\right)
\end{equation}
where $V_i \equiv V_{GR}$ for detuned scenario and $V_i \equiv V_{GW}$ for GW stabilized scenario. Now if we consider inflationary epoch at the RS brane with the on-brane radion field $\Phi_\pm$ serving as the slowly rolling spatially homogeneous inflaton field, then we can drop the kinetic piece in the action (\ref{80}) with the slow roll approximations. Hence the on-brane effective action looks similar to the Jordan frame $f(R)$ action in eq.(\ref{73}). So in inflationary epoch, the on-brane theory reduces to that of $f(R)$ gravity and the same (in)stability analysis works there. In sections \ref{Tuned-RS}, \ref{detuned-RS} and \ref{GW}, we have calculated the different effective potentials of the Einstein frame radion field. Using those Einstein frame scalar potentials, we can reconstruct the dual on-brane $f(R)$ theories at the inflationary epoch. This will be addressed in our future projects.\\

\noindent In addition to that, there are also several interesting extensions to our present work - RS with a time-dependent stabilizing field~\cite{61}, $f(R)$ braneworlds~\cite{63,Karmakar}, multi-brane extensions~\cite{65,66} and other higher dimensional modified gravity scenarios where the low-energy effective theory is of scalar-tensor type. Some of these will be addressed in our future works.

\section{Acknowledgment}
AK is supported through INSPIRE-SHE Scholarship by the Department of Science and Technology, Government of India. AK acknowledges helpful discussions with Soham Bhattacharyya and Gahan Chattopadhyay. The authors would also like to
thank the reviewers for helping to improve the manuscript.

\section*{Data Availability Statement}

Being a theoretical study, no experimental data is associated with this work.

\section*{Code Availability Statement}

Apart from plotting the figures, no major code is associated with this work. This shall be made available on reasonable request.

\appendix
\section{Radion field in Jordan and Einstein frames and the Coupling Integral}\label{A}

Recall from eq.(\ref{24}), (\ref{31}) and (\ref{36}) we have
\begin{equation}\label{104}
\alpha(\varphi_\pm) = - \sqrt{\frac{4\pi}{3 + 2\omega_\pm(\Phi_\pm)}}\;\;\;;\;\;\;\frac{d\varphi_\pm}{d\ln\Phi_\pm} = \sqrt{\frac{3 + 2\omega_\pm(\Phi_\pm)}{16\pi}} = \Phi_\pm \frac{d\varphi_\pm}{d\Phi_\pm}
\end{equation}
\begin{equation}\label{105}
\omega_+(\Phi_+) = \frac{3}{2}\frac{\Phi_+}{\gamma - \Phi_+}\;\;\;;\;\;\; \omega_-(\Phi_-) = -\frac{3}{2}\frac{\Phi_-}{\gamma + \Phi_-}
\end{equation}
We will first want to express the Einstein frame radion field $\varphi_\pm$ in terms of the BD scalar (or Jordan frame) $\Phi_\pm$ from eq.(\ref{104})
\begin{align}
\varphi_\pm = \int d\Phi_\pm \frac{1}{\Phi_\pm}\sqrt{\frac{3 + 2\omega_\pm(\Phi_\pm)}{16\pi}}
\end{align}
First we will do for the Planck brane
\begin{align}
\varphi_+ = \int d\Phi_+ \frac{1}{\Phi_+}\sqrt{\frac{3 + 2\omega_+(\Phi_+)}{16\pi}} &= \sqrt{\frac{3\gamma}{16\pi}}\int d\Phi_+ \frac{1}{\Phi_+\sqrt{\gamma - \Phi_+}} \nonumber \\
&= \sqrt{\frac{3}{16\pi}}\ln\left[\frac{\sqrt{\gamma} - \sqrt{\gamma - \Phi_+}}{\sqrt{\gamma} + \sqrt{\gamma - \Phi_+}}\right]
\end{align}
We can invert this relation to find $\Phi_+$ in terms of $\varphi_+$,
\begin{equation}\label{108}
    \Phi_+ = \gamma\;\text{sech}^2\left(\sqrt{\frac{4\pi}{3}}\varphi_+\right)
\end{equation}
This is the same expression as in eq.(\ref{a51}). Now we will do for the TeV brane
\begin{align}
\varphi_- = \int d\Phi_- \frac{1}{\Phi_-}\sqrt{\frac{3 + 2\omega_-(\Phi_-)}{16\pi}} &= \sqrt{\frac{3\gamma}{16\pi}}\int d\Phi_- \frac{1}{\Phi_-\sqrt{\gamma + \Phi_-}} \nonumber \\
&= -\sqrt{\frac{3}{16\pi}}\ln\left[\frac{\sqrt{\gamma + \Phi_-} + \sqrt{\gamma}}{\sqrt{\gamma + \Phi_-} - \sqrt{\gamma}}\right]
\end{align}
Again, we can invert this relation to find $\Phi_-$ in terms of $\varphi_-$,
\begin{equation}\label{110}
    \Phi_- = \gamma\;\text{csch}^2\left(\sqrt{\frac{4\pi}{3}}\varphi_- \right)
\end{equation}
This is also the same expression as in eq.(\ref{a61}). Finally we will do the integration of the matter coupling function $\alpha(\varphi_\pm)$ with respect to the Einstein frame radion field $\varphi_\pm$. Using eq.(\ref{104}) and (\ref{105}) we get
\begin{align}
\int d\varphi_\pm\;\alpha(\varphi_\pm) = \int d\Phi_\pm \frac{d\varphi_\pm}{d\Phi_\pm}\alpha(\Phi_\pm) &= -\int d\Phi_\pm \frac{1}{\Phi_\pm} \sqrt{\frac{3 + 2\omega_\pm(\Phi_\pm)}{16\pi}}\sqrt{\frac{4\pi}{3 + 2\omega_\pm(\Phi_\pm)}} \nonumber \\
&= -\frac{1}{2}\ln\Phi_\pm
\end{align}
Thus using eq.(\ref{108}) and (\ref{110}) we get
\begin{equation}
    \int d\varphi_+\;\alpha(\varphi_+) = -\frac{1}{2}\ln\left[\gamma\;\text{sech}^2\left(\sqrt{\frac{4\pi}{3}}\varphi_+\right)\right] \;\;\;\; ; \;\;\;\; \int d\varphi_-\;\alpha(\varphi_-) = -\frac{1}{2}\ln\left[\gamma\;\text{csch}^2\left(\sqrt{\frac{4\pi}{3}}\varphi_-\right)\right]
\end{equation}
which is the same as eq.(\ref{a39}).

\end{document}